\documentclass[12pt]{article}
\usepackage{a4}
\usepackage{amsfonts}
\usepackage{amssymb}
\usepackage{cite}
\usepackage{epsfig}
\usepackage{amsmath}
\usepackage{cite}
\author{}

\newcommand{\drawsquare}[2]{\hbox{%
\rule{#2pt}{#1pt}\hskip-#2pt
\rule{#1pt}{#2pt}\hskip-#1pt
\rule[#1pt]{#1pt}{#2pt}}\rule[#1pt]{#2pt}{#2pt}\hskip-#2pt
\rule{#2pt}{#1pt}}
\newcommand{\fund}{\raisebox{-.5pt}{\drawsquare{6.5}{0.4}}}
\newcommand{\Ysymm}{\raisebox{-.5pt}{\drawsquare{6.5}{0.4}}\hskip-0.4pt%
         \raisebox{-.5pt}{\drawsquare{6.5}{0.4}}}
\newcommand{\Yasymm}{\raisebox{-3.5pt}{\drawsquare{6.5}{0.4}}\hskip-6.9pt%
        \raisebox{3pt}{\drawsquare{6.5}{0.4}}}
\newcommand{\antifund}{\overline{\fund}}

\newcommand{\be}{\begin{equation}}
\newcommand{\ee}{\end{equation}}
\newcommand{\ba}{\begin{array}}
\newcommand{\ea}{\end{array}}
\newcommand{\bea}{\begin{eqnarray}}
\newcommand{\eea}{\end{eqnarray}}
\newcommand{\mbb}{\mathbb}

\newcommand{\ov}{\overline}
\def\IR{\relax{\rm I\kern-.18em R}}

\def\IP{\relax{\rm I\kern-.18em P}}

\def\inbar{\vrule height1.5ex width.4pt depth0pt}
\def\IC{\relax\,\hbox{$\inbar\kern-.3em{\rm C}$}}

\def\K3{{\bf K3}}

\def\ov{\overline}

\def\n2d{\cN_{V^*}^{\otimes 2}}

\def\IC{\mathbb{C}}

\def\IR{\mathbb{R}}

\def\IP{\mathbb{P}}

\def\cN{{\mathcal N}}

\def\to{\rightarrow}

\begin{document}

\title{
\begin{flushright} \vspace{-2cm}
{\small MPP-2006-119\\
 \small NSF-KITP-06-73\\
 \small UPR-1162-T\\
}
\end{flushright}
\vspace{1.8cm}
Spacetime Instanton Corrections in \\
4D String Vacua \\
{\large  ( - The Seesaw Mechanism for D-Brane Models - )}
}
\vspace{2.5cm}
\author{\small Ralph~Blumenhagen$^{1,3}$, Mirjam Cveti{\v c}$^{2,3}$  and  Timo Weigand$^{2}$}

\date{}

\maketitle

\begin{center}
\emph{$^{1         }$ Max-Planck-Institut f\"ur Physik, F\"ohringer Ring 6, \\
  80805 M\"unchen, Germany } \\
\vspace{0.1cm}
\emph{$^{2        }$ Department of Physics and Astronomy, University of Pennsylvania, \\
     Philadelphia, PA 19104-6396, USA } \\
\vspace{0.1cm}
\emph{$^{3        }$ Kavli Institute for Theoretical Physics,
University of California, \\
 Santa Barbara, CA 93106-4030, USA} \\
\vspace{0.2cm}

\tt{blumenha@mppmu.mpg.de, cvetic@cvetic.hep.upenn.edu, timo@sas.upenn.edu}
\vspace{0.1cm}
\end{center}
\vspace{1.0cm}

\begin{abstract}
\noindent We systematically investigate instanton corrections from wrapped
Euclidean D-branes to the matter field superpotential of various classes of
N=1 supersymmetric D-brane models in four dimensions. Both gauge invariance
and the counting  of fermionic zero modes provide  strong constraints on the allowed
non-perturbative  superpotential couplings. We outline how the complete
instanton computation boils down to the computation of open string disc
diagrams for boundary changing operators multiplied by a one-loop vacuum
diagram. For concreteness we  focus on E2-instanton effects in Type IIA vacua
with intersecting D6-branes, however the same structure emerges for Type IIB
and heterotic  vacua. The instantons wrapping rigid cycles can potentially
destabilise the vacuum or generate perturbatively absent matter couplings such
as proton decay operators, $\mu$-parameter or right-handed neutrino Majorana
mass terms. The latter allow the realization of the seesaw mechanism for MSSM
like intersecting D-brane models.

\end{abstract}

\thispagestyle{empty}
\clearpage

\tableofcontents

\section{Introduction}

Recent years have witnessed the emergence of a  new picture of the space of string vacua.
On the one hand, it is by now widely appreciated that the introduction of non-trivial background fluxes induces a computable scalar potential. The resulting plethora of stable minima of this potential has given rise to the idea of a landscape
of string vacua.
On the other hand, progress in string model building has taught us that also
the gauge sector of the perturbative four-dimensional string vacua including both heterotic
and D-brane constructions is much richer than anticipated.

As it stands, however, this picture is not complete
due to the existence of important non-perturbative effects in string theory,
which are well-known to have dramatic consequences. In particular, in the absence of any other
stabilising effects like fluxes they may  destabilise the entire string vacuum or generate
new couplings which have to be taken into account both for
the analysis of the scalar potential and for string phenomenology.
Attempts in the literature to confront  such issues in string-motivated constructions include the KKLT scenario,
where precisely these non-perturbative effects play a starring role, namely in  freezing all
moduli in a Type IIB orientifold compactification.

Let us stress that the non-perturbative effects are uniquely determined for
each concrete string vacuum - they cannot simply be chosen at our will to  add
suitable terms to the potential which  freeze  some of the moduli. One might
argue that for small string coupling they can be neglected as compared to the
perturbative or tree-level contributions. However, there do exist
important couplings which are only generated by instanton effects and
therefore constitute  the leading order terms which determine the physics. In
particular in addressing the  cosmological constant problem, instanton effects cannot be
neglected. Therefore, it is of utmost importance for a better understanding of
the string landscape to study these non-perturbative effects for all known
classes of four-dimensional string vacua. In the context of heterotic string
models this problem was taken seriously from the very beginning, but for
D-brane models it has apparently not received the full attention it deserves.

The aspects of instanton corrections  discussed
in the literature  include world-sheet instanton corrections for the
heterotic  string
\cite{Dine:1986zy,Dine:1987bq,Distler:1987ee,Witten:1999eg,Buchbinder:2002ic,Buchbinder:2002pr}
and Type IIA string \cite{Kachru:2000ih,Kachru:2000an,Aganagic:2000gs},
M2- and M5-brane instantons in M-theory
\cite{Becker:1995kb,Harvey:1999as,Belani:2006wx}, D-instanton effects in the D3-D(-1) system
\cite{Green:1997tv,Gutperle:1997iy,Billo:2002hm} in Type IIB string theory, inducing e.g. corrections to $R^4$
terms,  and
the effect of background fluxes on E2-instantons
\cite{Kashani-Poor:2005si}, to name but a few. A most prominent role alluded to already is played  by E3-instantons in Type IIB theory due to their
dependence on the K\"ahler moduli\cite{Witten:1996bn}.

It is the aim of this paper to systematically investigate spacetime instanton effects with special emphasis on
D-brane models, where for concreteness we start our investigations in the framework of Type IIA
intersecting D-branes (see \cite{Blumenhagen:2005mu} for a review).
Explicit examples of semi-realistic four-dimensional string vacua have been constructed in this framework on various
background geometries. Whereas in the framework of general Calabi-Yau manifolds the description of special Lagrangian (sLag) submanifolds wrapped by the D6-branes is a notoriously difficult problem,
it is among the virtues of toroidal orbifolds that they contain a large class of tractable supersymmetric three-cycles. In addition, the exactly known underlying conformal field
theory (CFT) allows for the computation of some of the perturbative low energy couplings  at string tree level.

A distinctive feature of intersecting brane worlds is the appearance of
$U(N_a)$ gauge factors on the stacks of D6-branes, whose abelian $U(1)$ part
is often anomalous. In this case, the Green-Schwarz mechanism effectively
converts them into global $U(1)$ symmetries in that they are, while broken as
a gauge symmetry, still respected by all perturbative correlation functions.
These global $U(1)$s are often a nuisance for string phenomenology as they
forbid some highly desirable charged matter couplings such as certain Yukawa
couplings for $SU(5)$ models or Majorana masses for right-handed neutrinos.

In this paper we address the question whether charged matter superpotential couplings
\bea
\label{ampl}
            W\simeq \prod_{i=1}^M  \Phi_i\, e^{-S_{\rm inst.}},
\eea
violating the global $U(1)$ symmetries,
can be generated non-perturbatively. In fact, the global $U(1)$s
and the associated gauged global shift symmetries of the
axions can be broken by Euclidean D2-brane instantons, called E2-instantons
in the sequel.
The possible appearance of a coupling of the form (\ref{ampl}) was first considered in \cite{Dine:1987bq} in the context of the heterotic string.
As already mentioned above, non-perturbative effects also play a crucial role in the context of moduli stabilisation.
As one of their most dramatic effects, they have the potential to destabilise the vacuum
completely or, often in combination with other ingredients like fluxes,
to drive it to an isolated minimum in the string landscape.

We outline the general structure of such non-perturbative contributions to the superpotential of the form (\ref{ampl}), relegating the explicit computation of the
instanton amplitudes to \cite{bcw}.
The key ingredients, analysed in section 2 and 3, are given by the issues of charge conservation and the counting of fermionic
zero modes.
The latter yield strong constraints on the possible couplings.
We also discuss how the instanton computation works
at the level of conformal field theory and find a
structure analogous to the one derived in more spacetime oriented
approaches to the problem. In addition we make  some comments
on multi-instanton corrections and instanton corrections to
the D-terms and the gauge kinetic function.

While for concreteness we will explain our methods in the setup of Type IIA D-brane models,
 we will point out in section 4 how
the same structure found in the Type IIA context emerges  for  Type IIB
orientifolds and for heterotic string models. In fact, our analysis  is
related, via a chain of dualities, to the investigation of world-sheet
instanton effects for the $Spin(32)/\mbb{Z}_2$ heterotic string
\cite{Witten:1999eg}. We will see that some of the more abstract notions
appearing in the heterotic computation have a clear  geometric and
conformal field theoretic interpretation for E2-instantons in Type IIA. Moreover,
our analysis clarifies that for dual Type IIB orientifolds, the naive uplifting
mechanism in the KKLT scenario by magnetised D7-branes does not work due to
the appearance of extra fermionic zero modes spoiling the generation of the
superpotential term for Euclidean D3-brane instantons. Only by multiplying the
exponential factor by appropriate charged matter fields the uplifting
might eventually work.

Eventually, section 5 analyses some E2-instanton generated effects of immediate phenomenological interest  for MSSM-like Type IIA intersecting D-brane models.
These involve  destabilisation mechanisms on the one hand and the possible appearance of perturbatively forbidden matter couplings on the other. The latter include  in principle dangerous dimension-four proton decay operators, but also hierarchically small $\mu$-terms and Majorana type masses for right-handed neutrinos.
\pagebreak

Let us summarise our main findings:
\begin{itemize}
\item{The global $U(1)$ charges formally carried (due to the Green-Schwarz
     mechanism)  by the exponential
     factor
      in the instanton amplitude have their microscopic origin  in
     extra fermionic zero modes living at the  intersections of the instanton
    with the spacetime filling D-branes present in the model.}
\item{
Instantons wrapping rigid supersymmetric cycles can generate charged
matter couplings in the superpotential, which can
potentially destabilise the vacuum or lead to  new effects
in the four-dimensional action which are absent in perturbation theory. In this case their contribution,
though exponentially suppressed, yields the leading order terms. }
\item{The technical treatment of such instanton amplitudes
    boils down to the computation of conformal field theory disc and one-loop
   diagrams with boundary changing  operators inserted
   along the boundary (see also \cite{Billo:2002hm}). These diagrams are multiplied by a non-vanishing one-loop vacuum
  diagram, in  which only the open string sectors preserving precisely two
    supercharges contribute. We present  the CFT analogue of the one-loop Pfaffian/determinant
    known to generally appear in instanton computations \cite{Witten:1999eg}. }
\item{By T-duality, E2-instantons in Type IIA orientifolds are mapped
to  E1/E5- and E3-instantons in Type IIB orientifolds. S-duality maps
the E1-instantons to world-sheet instantons for the heterotic string.
Many of the features of the latter have a corresponding  geometric or open string interpretation
in the Type IIA framework.}
\item{Instantons  with appropriate intersections
with the two D-branes supporting the right-handed
neutrinos on their intersection can give rise to
Majorana mass-terms. The scale of these masses
is
\bea
           M_m=x\, M_s\, e^{-{2\pi\over \ell_s^3\, g_s} {\rm Vol}_{E2} }
\eea
with $x$ expected to be of order $O(1)$.
These  can easily generate a hierarchy between the string and
the Majorana mass scale.
}
\end{itemize}

\vspace{10pt}
\noindent {\bf Note:} We thank the authors of  \cite{Shamit} and \cite{dieter} for pointing out to us their work (see also \cite{McGreevy}), which likewise deals with instanton corrections in Type II D-brane models.
\vspace{10pt}

\section{The GS mechanism and global $U(1)$s}

As our prototype class,
we consider the Type IIA string compactified on a Calabi-Yau (CY) threefold
${\cal X}$ modded out by an orientifold projection $\Omega\overline{\sigma}
(-1)^{F_L}$. The world-sheet parity transformation $\Omega$ is combined
with an anti-holomorphic involution of ${\cal X}$.

Following mainly \cite{Blumenhagen:2002wn},
on the threefold we introduce in the usual way a symplectic
basis $(A_I, B^I)$, $I=0,1,\ldots , h_{21}$  of homological three-cycles with
the topological intersection  numbers
\bea
             A_I\circ B^J = \delta_{I}^J.
\eea
Upon introducing the corresponding  three-forms satisfying
\bea
          \int_{A_I} \alpha^J =\delta_{IJ}, \quad \int_{B^I}
          \beta_J=-\delta_{IJ},
\eea
 it follows that the Poincar\'e dual of the three-cycles $(A_I,B^I)$ are
the three-forms $(\beta_I,\alpha^I)$.

Let us denote by $\Pi_{O6}$ the homology class of the
orientifold plane, which is the homology class of the fixed point locus
of the anti-holomorphic involution $\overline{\sigma}$. The latter is
a special Lagrangian three-cycle in ${\cal X}$.
For tadpole cancellation, one introduces stacks of ${\cal N}_a$ D6-branes
wrapping special Lagrangian three-cycles of homology class
$\Pi_a$. Their orientifold images are denoted by $\Pi'_a$.
Expanding such three-cycles into the symplectic basis
\bea
   \Pi_a= M^I_{a}\, A_I + N_{a,I}\, B^I
\eea
with $M^I, N_I\in \mbb{Z}$,
one obtains for the topological intersection number between two
three-cycles $\Pi_a, \Pi_b$
\bea
    \Pi_a\circ \Pi_b= M^I_{a}\, N_{b,I}-N_{a,I}\, M^I_{b}.
\eea
The tadpole cancellation condition for an intersecting
D-brane model of this type reads
\bea
    \sum_{a=1}^K  {\cal N}_a \left( \Pi_a + \Pi'_a\right) = 4\, \Pi_{O6}.
\eea
For the generic  gauge group $G=\prod_{a=1}^K U({\cal N}_a)$
the chiral massless spectrum  is given in table \ref{tcs}.

\begin{table}[h]
\centering
\label{tcs}
\begin{tabular}{|c|c|c|}
\hline
Non-abelian Reps. & $U(1)$ charges &    Multiplicity \\
\hline \hline
$\Yasymm_a$
 & $(2_a)$ & ${1\over 2}\left(\Pi'_a\circ \Pi_a+\Pi_{{\rm O}6}
\circ \Pi_a\right)$  \\
$\Ysymm_a$
   & $(2_a)$ &   ${1\over 2}\left(\Pi'_a\circ \Pi_a-\Pi_{{\rm O}6} \circ \Pi_a\right)$   \\
$(\antifund_a,\fund_b)$
 & $(-1_a,1_b)$ & $\Pi_a\circ \Pi_{b}$   \\
 $(\fund_a,\fund_b)$
 & $(1_a,1_b)$ & $\Pi'_a\circ \Pi_{b}$
\\
\hline
\end{tabular}
\caption{Chiral spectrum for intersecting D6-branes} 
\end{table}

Note that this is just the chiral spectrum. For the non-chiral
part one has to examine the intersections more carefully.
For this purpose, one introduces the physical intersection number
between two branes, which splits into the number of positive
and negative intersections
\bea
          \Pi_a\cap \Pi_b =[\Pi_a\cap \Pi_b]^+ + [\Pi_a\cap \Pi_b]^-.
\eea
The topological intersection number is then given by
\bea
             \Pi_a \circ \Pi_b =[\Pi_a\cap \Pi_b]^+ - [\Pi_a\cap \Pi_b]^-.
\eea
In general, the   abelian subgroup
$U(1)_a\subset U({\cal N}_a)$  is anomalous. It is known that these anomalies
are cancelled by a generalised Green-Schwarz mechanism involving
the four-dimensional axions coming from dimensional reduction
of the ten-dimensional R-R three- and five-form \cite{Aldazabal:2000dg}.
If we expand $C^{(5)}$ in the symplectic basis
\bea
          C^{(5)}=\ell_s^3\left( C^{(2)}_I\, \alpha^I - D^{(2),I}\, \beta_I
   \right),
\eea
dimensional reduction of the Chern-Simons couplings in the WZW action
gives rise to the following mixing term between the four-dimensional axions
and the $U(1)$ gauge fields $F_a = dA_a$
\bea
     S_{mix}= {{\cal N}_a\over \ell_s^2}\,  \int_{\IR^{1,3}} \left(  (M^I_{a}-{M'}^I_{a})\,
     F_a\wedge C^{(2)}_I
                       + (N_{a,I}-N'_{a,I})\, F_a\wedge D^{(2),I}  \right).
\eea
The minus sign in front of the contribution from the image branes
come from the fact that under $\Omega$ the $U(1)_a$ field strength transforms
as $F_a\to -F_a$.
Expanding similarly the Hodge dual three-form $C^{(3)}$
\bea
\label{cthreeexp}
          C^{(3)}=\ell_s^3\left( C^{(0)}_I\, \alpha^I - D^{(0),I}\, \beta_I
   \right),
\eea
one notices that in four dimensions $(C^{(2)}_I, D^{(2),I})$
is Hodge dual to $(-D^{(0),I}, C^{(0)}_I)$.
The Green-Schwarz mechanism now implies that under a $U(1)_a$
gauge transformation
\bea
            A_{a,\mu}\to A_{a,\mu}+ \partial_{\mu} \Lambda_a
\eea
the axions $(C^{(0)}_I, D^{(0),I})$ transform as
\bea
\label{gaugetrafoa}
          C^{(0)}_I \to  C^{(0)}_I + Q^a_I \, \Lambda_a, \quad\quad
          D^{(0),I} \to  D^{(0),I} + P^{a,I} \, \Lambda_a
\eea
with
\bea
\label{gaugetrafob}
     Q^a_I={{\cal N}_a\over 2\pi}\left(  N_{a,I}-N'_{a,I}\right), \quad\quad
     P^{a,I}=-{{\cal N}_a\over 2\pi}\left(  M^I_{a}-{M'}^I_{a}\right).
\eea
Therefore, the  $U(1)$ gauge potentials and axions which remain massless have to lie in the
left and right kernel, respectively, of the
$(K,2h_{21}+2)$ matrix ${\cal Q}=(Q,P)$.
The $U(1)$ symmetries corresponding to the massive gauge potentials still survive as perturbative global
symmetries in that all perturbative correlation functions respect them.

Since some of the axions become the longitudinal modes of the massive
$U(1)$ gauge fields, also their superpartners must become massive
for supersymmetry preserving configurations. These mass terms for
the corresponding complex structure moduli arise from the
generated Fayet-Iliopoulos terms
\bea
\label{fiterm}
      \xi_a\simeq {\rm {Arg}} \int_{\Pi_a}  \Omega_3.
\eea

From dimensional reduction of the DBI action one can
deduce the $SU({\cal N}_a)$ gauge kinetic functions to be given by\footnote{For supersymmetry, $\xi_a=0$ and clearly $ \int_{\Pi_a} 
     \Re (\Omega_3)= \big| \int_{\Pi_a} \Omega_3  \big| $.}
\bea
\label{gaugecoup}
     f_a={1\over (2\pi)\, \ell_s^3}\left[ {1\over g_s}  \int_{\Pi_a} 
      \Re (\Omega_3)  \, -i \int_{\Pi_a} C^{(3)} \right].
\eea
With the normalisation ${\rm tr}(T_a\, T_b)={1\over 2}\delta_{ab}$,
one obtains for the physical non-abelian gauge coupling
\cite{Klebanov:2003my,Blumenhagen:2003jy}
\bea
    {4\pi\over g_a^2}={1\over g_s \ell_s^3}\, {\rm Vol}_{D6_a}.
\eea

\section{E2-instantons}
\label{E2_gen}

The question we address in this section concerns the
effects of spacetime instantons on the type of models just described.
Such non-perturbative effects are generated in principle
by wrapped E0, E2 and E4-branes. However, since a Calabi-Yau manifold
does not have any continuous one and five-cycles, the only relevant\footnote{We will not explore here the
possibility of E4-instantons wrapping homologically trivial, but fluxed five-cycles being the Euclidean version
of the coisotropic eight-branes proposed in the intersecting brane context recently \cite{Font:2006na}.
Their effect, if relevant, is expected to be qualitatively of the same structure as the one of E2-instantons
investigated in this article.}
instantons are E2-branes wrapped on three-cycles of the CY threefold.
Additional non-perturbative effects arise from  world-sheet instantons and in principle
there could also be
NS5-brane instantons wrapping the entire CY. The latter would be magnetically charged under
the NS-NS two-form $B_2$ with both legs along  $\IR^{(1,3)}$ and are therefore
modded out by the orientifold projection.

For ease of presentation our detailed discussion will  be restricted to single-instanton effects, whereas we will indicate the relevant modifications required for the multi-instanton case in subsection \ref{multi}.

\subsection{Global $U(1)$ charges}

The E2-branes couple to the
 R-R form $C^{(3)}$ and therefore to the axions whose shift symmetries
have been gauged for the cancellation of the $U(1)$ anomalies. Consequently, such an instanton transforms
non-trivially under abelian gauge symmetries.

Let us consider a single E2-brane wrapping a homological three-cycle
\bea
    \Xi=E^I\, A_I + F_I\, B^I.
\eea
The orientifold action requires the introduction of an E2-brane wrapping also
the $\Omega\overline{\sigma} (-1)^{F_L}$ image cycle $\Xi'$.
Since we are interested in contributions to the ${\cal N}=1$ superpotential,
the three-cycle should preserve supersymmetry, i.e. it should
be an sLag cycle. Even more, in the internal CY space this cycle should
preserve the same supersymmetry as the orientifold plane
and the D6-branes do.

Since the E2-brane is localised in the flat
four dimensions spanned by the D6-branes the whole
configuration breaks one-half of the supersymmetry so that
two supersymmetries  are conserved and two are broken.
It follows that the E2-brane world-volume accommodates at least
the two fermionic zero modes $\theta_i$ corresponding to the broken supersymmetries, as required for a
 contribution to the superpotential.

Any correlation function in this single-instanton background
contains the classical factor
\bea
\label{expfac}
             e^{-S_{E2}}=\exp\left[ -{2\pi\over \ell_s^3}
           \left( {1\over g_s}\int_{\Xi} \Re(\Omega_3) - i \int_{\Xi} C^{(3)}
         \right) \right],
\eea
which depends exponentially on the complex structure moduli $U$.
In fact, the emergence of this factor follows directly from the CFT formalism to be presented in section
 \ref{calc} (see also \cite{Polchinski:1994fq}).

The expansion of $C^{(3)}$ in equ. (\ref{cthreeexp}) and the gauged shift symmetries
(\ref{gaugetrafoa}) allow us to compute  the transformation properties of this
instanton factor under the $U(1)_a$ gauge symmetries to be
\bea
       e^{-S_{E2}}\to  e^{i\, Q_a(E2)\,\Lambda_a}  \,\, e^{-S_{E2}}
\eea
with
\bea
\label{chargee}
          Q_a(E2)={\cal N}_a\,\, \Xi\circ (\Pi_a - \Pi'_a) .
\eea
Therefore, the $U(1)$ charge of this term is 
given by the topological
intersection number of the three-cycle wrapped by the E2-instanton  and
the three-cycles wrapped by the D6-branes. The relative minus sign
between $\Pi_a$ and its image brane $\Pi'_a$ is just right
to take care of the fact that for invariant D6-branes there is no
$U(1)_a$ gauge field.

If any of the charges $Q_a(E2)$ is non-vanishing, it is clear
that a purely exponential superpotential of the form $W=c\cdot \exp ( -S_{E2})$ cannot be generated due to
gauge invariance. Only if the exponential factor gets multiplied
by a suitable combination of matter superfields also charged under $U(1)_a$ can
a  superpotential
\bea
\label{superproduct}
                 W=  \prod_i \Phi_i\   e^{-S_{E2}}
\eea
be gauge invariant. Here appropriate contractions
of indices are understood  so that the matter field product is
invariant under all the non-abelian gauge symmetries.

Note that while the product (\ref{superproduct}) is still invariant under the $U(1)_a$ global symmetries, of
course, the induced coupling of the superfields $\Phi_i$ clearly violates these abelian symmetries - after
all we have to evaluate $e^{-S_{E2}}$ at its classical value which basically involves just the volume of the
three-cycle $\Xi$. In this sense the $U(1)_a$ global symmetries
in the four-dimensional effective action are broken by
E2-instantons.

So far we have merely investigated selection rules
resulting from the abelian gauge symmetries for the generation
of a non-perturbative superpotential. However, these are merely  necessary conditions
for the appearance of an instanton induced coupling. To determine whether these terms are really present
requires a genuine instanton computation in which further issues such as the number of fermionic zero modes
play an important and restrictive role.

\subsection{Zero modes}

Of prime importance for the  computation of an $n$-point function
in an instanton background are the
fermionic zero modes. As is well-known and will be reviewed in detail momentarily, these have  to be integrated over and
therefore they have to be absorbed by the vertex operators appearing in the $n$-point correlator.
In the present context, it is important to distinguish between two very different kinds of
fermionic zero modes depending on whether or not they are charged under the abelian gauge groups.

We have already pointed out the presence of the usual two fermionic zero modes $\theta_i$
corresponding to the broken supersymmetries  of the half-BPS instanton
which enable it to contribute to the superpotential.
These zero modes arise from the Ramond sector of an open string with both ends attached to
the E2-instanton.
From the CFT point of view they are described by the vertex operator (in the $(-1/2)$-ghost picture)
\bea
     V^{-{1\over 2}}_{\Theta}(z)=\theta_i\,\, e^{-{\varphi(z)\over 2}}\,\, \Sigma^{E2,E2}_{3/8}(z)\,\,
                 S^i(z),
\eea
where $\theta_i$, $i=1,2$  is the polarisation and $S^i$ denotes the 4D spin field of $SO(1,3)$. This
is a Weyl spinor  of conformal dimension $h={1/4}$. The twist field $\Sigma^{E2,E2}_{3/8}$
is essentially the spectral flow operator of the ${\cal N}=2$ superconformal field theory
describing the internal CY manifold. Since the E2-brane wraps a three-cycle on the CY manifold,
the 4D spacetime is transversal so that there appears  no 4D momentum factor  in the
vertex operator.

If the three-cycle $\Xi$  has deformations counted
by $b_1(\Xi)$, then there will be extra fermionic zero modes arising from open strings from the
E2 to itself. Another source of zero-modes are the
intersections of $\Xi$ with its image brane $\Xi'$.
Since on a single brane there are no fields transforming in the
anti-symmetric representation of $U(1)_{E2}$ and the $\Omega$ projection changes
sign for the E2-branes relative to the D6-branes, the condition for the absence
of extra uncharged zero modes reads
\bea
\label{nosyms}
         [\Xi'\cap \Xi]^{\pm} = -[\Pi_{O6}\cap \Xi]^{\pm} = 0
\eea
for single instantons.

In the sequel we call the
zero modes of the type just described uncharged as they carry no charge under any of the gauge
symmetries localised on the D6-branes.
In order to lead to a non-vanishing F-term, all these extra uncharged zero modes
 have to be absorbed by additional closed or open string fields. \footnote{In this case, we also have to integrate over the bosonic components of the $b_1(\Xi)$ superfields describing the instanton moduli. We expect, however, that such configurations do not contribute to the superpotential, but rather to higher derivative F-terms, see \cite{Beasley:2005iu} for the analogous effect in the context of heterotic (0,2) models.}
As a result, if we want to generate a contribution to the
superpotential of only the charged matter fields, we require
these uncharged zero modes to be absent. In particular, this means
that the sLag cycle $\Xi$ needs to be rigid. The construction of such rigid cycles in toroidal orbifolds has been explored in \cite{Blumenhagen:2005tn,Blumenhagen:2006ab}.

In view of the discussion of the previous section, one might suspect that the formal
$U(1)_a$ charges (\ref{chargee})
of the instanton have a microscopic origin in the zero mode structure
on the world-volume of the E2-brane.
Indeed, at the intersection of E2 with a stack of D6-branes, there do  appear
extra charged states transforming in bi-fundamental representations.
To determine these states, we first note that generically for two  intersecting
D6-branes at non-trivial angles the Neveu-Schwarz and Ramond physical ground state energies for open strings stretching between
them are $E_{NS}>-{1\over 2}$ and $E_{R}=0$ \cite{Berkooz:1996km}. Moreover, at a supersymmetric
 intersection there lives
one 4D chiral superfield, i.e. two real bosonic and two fermionic on-shell
degrees of freedom.
Now, if one exchanges one of the D6-branes by an E2-brane
wrapping the same sLag three-cycle, the four flat Neumann-Neumann boundary conditions
for the open string become Neumann-Dirichlet boundary conditions. This has three
important effects for the open string partition function.
First, as is known from $\#_{ND}=4$ systems, the GSO projection changes sign
with respect to the GSO projection in the D6-D6 intersection. 
Second, the ground energies are now $E_{NS}+{1\over 2}>0$ and $E_R=0$ so that there
are no massless bosons at a non-trivial  intersection. Third, the number
of massless fermions is halved, so that  one obtains only one real or one-half
complex fermionic on-shell degree of freedom \footnote{As we will see in section \ref{sechet},
these configurations are mapped by a chain of dualities to world-sheet instantons
for $(0,2)$ heterotic string models. Here these $1/2$ complex fermionic zero modes
are the familiar left-moving zero modes discussed in \cite{Distler:1987ee,Witten:1999eg}.
These  sit in Fermi supermultiplets of the
$(0,2)$ supersymmetry, which indeed  do not contain any dynamical
bosonic superpartners.}. 

As is well-familiar from the intersecting brane context, for specific choices of the three-cycles wrapped by the D6 and E2-branes,
 one gets further extended supersymmetry
  at the intersection so that, as in \cite{Ganor:1996pe},
  a pair of fermionic zero modes
$(\lambda_a\overline{\lambda}_a)$ appears. Only if the E2-instanton
wraps the same three-cycle as the D6-brane or at special limiting points
in  the complex structure moduli space, also  complex bosonic zero
modes show up. It will  become clear, however, that these situations are of minor relevance for the computation of the superpotential terms we are primarily interested in.

Let us call the extra fermionic zero modes arising from open strings stretched
between the E2-brane and a stack of ${\cal N}_a$  D6-branes charged zero modes. The corresponding
Ramond sector open string vertex operators are of the form
\bea
          V^{-{1\over 2}}_{\Lambda^i_{a,I}}(z)=\lambda^i_{a,I}\, e^{-{\varphi(z)\over 2}}\,\, \Sigma^{a,E2}_{h=3/8}(z)\,\,
                             \sigma_{h=1/4}(z),
\eea
where $I=1,\ldots,[\Xi\cap \Pi_a]^+$ runs over the number of intersection points and
$i=1,\ldots,{\cal N}_a$ is the Chan-Paton (CP) index.
Here $\Sigma^{a,E2}_{h=3/8}$ denotes a spin field in the R-sector depending on the internal intersection
angles between the E2-brane and the D6-branes and is essentially the same operator
as it appears for vertex operators describing the fermionic zero modes for
two intersecting D6-branes \cite{Cvetic:2003ch,Cvetic:2006iz}. The 4D spin field $\sigma_{h=1/4}$ arises from the
twisted 4D world-sheet bosons carrying half-integer modes.
Note that also these zero modes carry no momentum along the flat 4D directions.

The total number of such charged fermionic zero modes is displayed  in Table \ref{tablezero}.

\begin{table}[ht]
\centering
\begin{tabular}{|c|c|c|}
\hline
zero modes&  Reps. & number   \\
\hline \hline
$\lambda_{a,I}$ &  $(-1_E,\fund_a)$   & $I=1,\dots, [\Xi\cap \Pi_a]^+$    \\
$\overline{\lambda}_{a,I}$ &  $(1_E,\antifund_a)$  & $I=1,\dots, [\Xi\cap \Pi_a]^-$    \\
\hline
$\lambda_{a',I}$ &  $(-1_E,\antifund_a)$ & $I=1,\dots, [\Xi\cap \Pi'_a]^+$    \\
$\overline{\lambda}_{a',I}$    &  $(1_E,\fund_a)$   & $I=1,\dots,[\Xi\cap \Pi'_a]^-$    \\
\hline
\end{tabular}
\caption{zero modes on E2, D6 intersections
\label{tablezero} } 
\end{table}

From the table it is clear that the total charge of all
the fermionic zero modes on the intersection of $\Xi$ and $(\Pi_a+\Pi'_a)$
is precisely $Q_a(E2)={\cal N}_a\, \Xi\circ (\Pi_a-\Pi'_a)$.

To conclude, the microscopic origin of the $U(1)_a$ charges
of the instanton factor is the appearance of extra
fermionic zero modes living on the intersection of the instanton
with the D6-branes. Therefore they are charged under
the gauge symmetry living on the D6-branes.

\subsection{Outline of instanton calculus}
\label{calc}

In this section, we outline how one technically  proceeds
in  computing the instanton corrections to the superpotential with special
focus on charged matter couplings.
As both the D6-branes and the E2-instantons have a CFT description
in terms of open strings, the computation can
be carried out in the CFT framework.
Of course, since we are still lacking  a string field theory framework,
one cannot derive the rules for instanton computations
from first principles but has to guess them
by invoking analogies with what we experience in field theory (see however \cite{Giddings:2005ff} for an account of non-perturbative effects directly in ten dimensions).
In the following we intend to focus on the general structure of such a computation,
making extra combinatorical factors and normalisations explicit only where they are essential.
Moreover, in order to avoid overloading the notation with too many indices, we suppress the intersection indices $I$ and the CP-indices $i$.
It is understood that all expressions we write down  have also to include these
labels in a consistent manner.

Our aim is to derive the contribution of an E2-instanton
to the  superpotential involving powers of charged chiral superfields
\bea
\label{chirsuper}
            W\simeq  \prod_{i=1}^M  \Phi_{a_i,b_i}\, e^{-S_{E2}},
\eea
where the $\Phi_{a_i,b_i}=\phi_{a_i,b_i}+\theta \psi_{a_i,b_i}$ are superfields localised at the
intersection of the D6-brane $\Pi_{a_i}$ with the D6-brane $\Pi_{b_i}$.
If a D6-brane has open string moduli, then this includes the case $a_i=b_i$. To make
this distinction more transparent, we will
call these latter superfields $\Delta_{a_i}=\Phi_{a_i,a_i}$ with bosonic components $\delta_a$ in the sequel.

The vertex operators for the bosons in the $(-1)$-ghost picture are  given by
\bea
       V^{-1}_{\phi_{a,b}}(z)=e^{-\varphi (z)}\,\, \Sigma^{ab}_{h=\frac{1}{2}}(z)\,\, e^{i p_\mu X^\mu (z)},
\eea
where $\Sigma^{ab}_{h=\frac{1}{2}}$ is a twist operator of conformal dimension $h=\frac{1}{2}$
depending on the intersection angles between the two D6-branes in the internal
CY directions (see \cite{Cvetic:2006iz} for the explicit form of these states for toroidal
models). The corresponding vertex operators for the fermions in the
$(-1/2)$-ghost picture are
\bea
      V^{-{1\over 2}}_{ \psi_{a,b}}(z)=\eta_{\alpha}\,
      e^{-{\varphi(z)\over 2}}\,\, \Sigma^{ab}_{h={3\over 8}}(z)\,\, S^{\alpha}(z)\, \,e^{i p_\mu X^\mu(z)},
\eea
where $S^\alpha$ is a spin field of $SO(1,3)$ of conformal dimension $h=1/4$.
For a D6-brane modulus or rather its bosonic component the corresponding boundary deformation vertex
 operator in the
$(0)$-ghost picture reads
\bea
  V^{0}_{\delta_{a}}(z)= \Sigma^{aa}_{h={1}}(z)\,\, e^{i p_\mu X^\mu (z)}.
\eea
Note again that extra intersection and CP-labels have been suppressed.
For ease of notation we will not
take care of the  image branes $\Pi'$ explicitly, but it
is understood that some of the matter fields in (\ref{chirsuper})
can also arise from intersections with the $\Pi'$.

\noindent A coupling of the type (\ref{chirsuper}) is detected by computing an appropriate
matter field correlation function in the E2-instanton background like
\bea
\label{instam}
       \langle \phi_{a_1,b_1}(p_1)\cdot\ldots  \cdot\phi_{a_{M-2},b_{M-2}}(p_{M-2})
     \cdot \psi_{a_{M-1},b_{M-1}}(p_{M-1})\cdot \psi_{a_{M},b_{M}}(p_{M})
\rangle_{E2-{\rm inst}}.
\eea
In the amplitude we have to integrate over all fermionic and bosonic
zero modes of the instanton, thus essentially taking into account all possible instanton configurations.
Since the correlation
function (\ref{instam}) we are probing for does not include any uncharged fields besides the two
broken supersymmetries, there may
be no uncharged fermionic zero modes other than $\theta_i$ present. Integration over them would make
the amplitude vanish.
As we explained already, this requires that the E2-brane
 wrap a rigid sLag three-cycle $\Xi$ and that there appear no
extra massless states on the intersection of $\Xi$ with $\Xi'$, i.e.
$\Xi'\cap \Xi=-\Pi_{O6}\cap \Xi=0$.

On the other hand, we also have to integrate over all   charged fermionic zero modes,
$\lambda_{a,I}$ and $\overline{\lambda}_{a,I}$
from the intersection of the E2-instanton  with the D6-branes. By the same argument as above, for an
 amplitude to be non-vanishing it has to involve each of these charged fermionic zero modes precisely
once.
We therefore propose that the complete instanton amplitude (\ref{instam})
is given by  the combinatorics of all possible non-vanishing CFT correlators
involving the E2-boundary
\bea
\label{instprop}
&&   \int d^4 x\, d^2\theta\,  \sum_{\rm conf.} \,
     {\textstyle
      \prod_{a} \left(\prod_{I=1}^{ [\Xi\cap
             \Pi_a]^+}  d\lambda_{a,I}\right)\,
               \left( \prod_{I=1}^{ [\Xi\cap
             \Pi_a]^-}  d\overline{\lambda}_{a,I}\right) } \nonumber \\
      &&\phantom{aaaaaaaaaaa}    \prod_k
\langle \Phi^k_{a_{k_1},b_{k_1}}\cdot \ldots \cdot \Phi^k_{a_{k_r},b_{k_r}}
          \rangle^{g_k}_{\prod \lambda_k}.
\eea
Here the sum over all configurations means that we sum over all
possible ways of assembling  the matter fields into sets (labelled
by $k$), for which
in the integrand one computes the open string CFT correlators of genus
$g_k$ with all
possible insertions of charged fermionic zero modes. Note that in the CFT amplitudes
one integrates over the world-sheet positions  of the CFT  vertex operators so
that the correlator only depends on the spacetime polarisation of the fields.

For getting the 1PI contribution we do not allow any disc correlator
without any zero modes attached, i.e. those discs without any
boundary on the E2-instanton. Even though these products
of CFT disc correlators with fermionic zero modes attached are
1PI reducible from the world-sheet point of view,
they are nevertheless 1PI irreducible from the spacetime point of view, as we integrate
over the zero modes.

Recall that for precisely two of the $\Phi_{a_i,b_i}$  we have to
choose the fermionic components $\psi_{a_i,b_i}$ and for the remaining ones
the bosonic components  $\phi_{a_i,b_i}$. This is coupled to the question where we insert the vertex
operators of the $\theta_i$, as described below. To keep the notation as simple as
possible, we will not make this issue explicit in the following formulas.

The evaluation of (\ref{instprop}) seems to be a daunting combinatorial exercise, but fortunately
most configurations do not contribute to the superpotential.
Of prime importance is the standard fact that the superpotential has to be a holomorphic function of the chiral
superfields. As is well-known, this implies that the only dependence on the string coupling
$g_s=e^{\phi}$ is via an exponential factor of the type (\ref{expfac})
involving also the axionic fields in the complex structure superfields.
Any other dependence
on these axions and therefore on $g_s$ can be excluded due to the axionic shift symmetries ($U(1)$ charges) \cite{Dine:1986zy,Dine:1987bq}. Note that in the CFT the matter fields are canonically normalised so
that the higher order corrections in $g_s$ reflect higher order
corrections to the K\"ahler potential.

To conclude, counting of factors of $g_s$ together with the need to insert all fermionic zero modes
including the two uncharged $\theta_i$ gives the terms which can, if non-zero, appear in the
superpotential. For this purpose, it is crucial to keep in mind the following normalisation factors
occurring:

\begin{itemize}
\item
Each disc diagram carries
an overall normalisation factor proportional to $ g_s^{-1}$.
Each annulus or M\"obius diagram comes with an additional factor of $g_s$.
From the counting of $g_s$ it is therefore  clear that  only world-sheets with
the topology of a disc or of an annulus or M\"obius strip, respectively,
can contribute to the superpotential. Moreover, at least one boundary
has to be the E2-instanton.

\item The vertex operators for the fields arising from D6-brane intersections
    do not carry any $g_s$ normalisation, as otherwise they would decouple
    in the weak coupling limit $g_s\to 0$.

\item
On the contrary, the charged fermionic zero modes can in principle
carry  a normalisation involving  a positive power of $g_s$,
as in the weak coupling limit the instanton decouples.
In fact, if we do not assign  any scaling to these zero modes, all disc
amplitudes would be of order $g_s^{-1}$ and could therefore not
appear in the superpotential. The minimal attachment  of fermionic zero
modes involves two per disc, which we take as a strong indication  that we should
assign a factor $\sqrt{g_s}$ to each  vertex operator $\Lambda_a$
\footnote{In the paper \cite{Billo:2002hm}, the authors came to exactly the same
conclusion, requiring that in the D3-D(-1) brane system
the zero modes localised on the intersection of these two branes
should  be identified with the gauge instanton moduli in the ADHM construction.
}.

As an immediate consequence, as far as the superpotential is concerned, these fermionic zero modes
can only contribute to the disc amplitudes, whereas
the 1-loop amplitudes have to be uncharged.
\end{itemize}

\vspace{20pt}

\noindent
\underline{Disc amplitudes} \\
\vspace{10pt}

\noindent Let us begin with the possible disc amplitudes which may contribute to the superpotential.
First, observe that one can always factor off the term $(\langle 1 \rangle_{disc})^n$ for all powers
$n$, as explained in \cite{Polchinski:1994fq,Billo:2002hm}. We have to sum over all such contributions, taking into account the factor $\frac{1}{n!}$
from the combinatorics of boundaries. This  leads to the familiar $\exp(-S_{E2})$
factor in the instanton amplitude.

\begin{figure}[h]
\begin{center}
\epsfbox{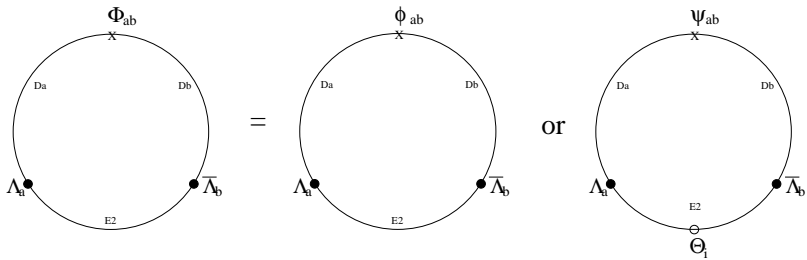} \caption{Standard disc tadpole} \label{figa}
\end{center}
\end{figure}

\noindent
The action of the E2-instanton depends holomorphically on the $\Omega\overline{\sigma} (-1)^{F_L}$
invariant complex structure moduli $U$. The non-renormalisation theorem of \cite{Brunner:1999jq}
for any polynomial
dependence of A-type disc diagrams on the complex structure moduli still applies,
reflecting the fact that the imaginary parts of the $U$ are axionic states with a shift symmetry.

Let us now turn to the non-trivial matter insertions, assuming for the time being that the $D6$-branes
 are rigid. The basic building block of order $(g_s)^0$
involves just one factor of $\Phi_{a_i,b_i}$ per disc.
\noindent
Depending on the number of $\theta$-modes inserted, we have the following options:
As shown in figure \ref{figa}, the simplest process is described by a disc diagram
with three boundary changing operators, the bosonic vertex operator corresponding to  $\phi_{a_i,b_i}$
being in the (-1) picture and the two fermionic vertex operators for $\lambda_{a_i}$ and
$\lambda_{b_i}$ in the (-1/2)-picture. With the above scaling of the involved vertex operators and the
normalisation of the disc amplitude, this amplitude is indeed of the order $(g_s)^0$, as required for
terms in the superpotential. If we are dealing with one of the fermionic modes
$\psi_{a_i,b_i}$, the disc also has to involve the vertex for one of the two $\theta_i$ in order that
the ghost number of all vertex operators can add up to 2, i.e. all fermions
can be chosen in the (-1/2)-picture.

Alternatively, the two $\theta_i$ modes could of course be absorbed in a disc involving the auxiliary
field $F_{a_i,b_i}$ and two charged fermion modes $\lambda$. This is what happens, for example, when
we probe for tadpoles of $\Phi_{a_i,b_i}$. Note that due to the $Spin(1,3)$ spinors present in the vertex operators of $\theta_i$ these are indeed the only ways to contract them  with the vertex operators of the component fields of $\Phi_{ab}$. This is the well-known CFT explanation why half-BPS instantons can only contribute to the superpotential.

\noindent
All these diagrams
can in principle be computed using conformal field theory techniques
and will subsequently be  denoted as
$\langle \Phi_{a_i,b_i}\rangle_{\lambda_{a_i},\lambda_{b_i}}$, where a consistent
assignment of intersection and CP-indices as well as
 an appropriate insertion of $\theta$-modes and the corresponding choice of components of
$\Phi_{a_i,b_i}$ is understood.

\begin{figure}[h]
\begin{center}
\hbox{ \hskip 2cm \epsfbox{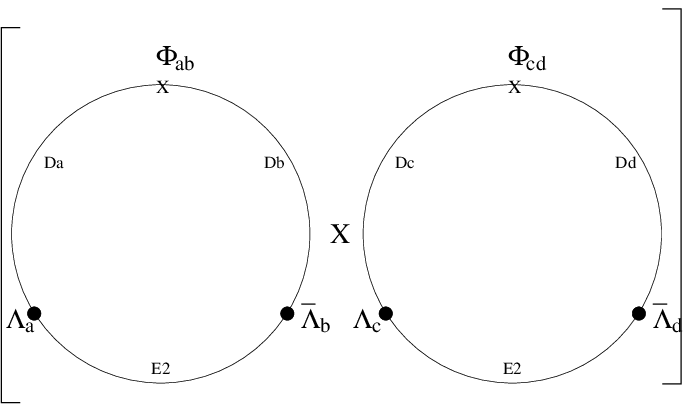}\hskip 1cm   \epsfbox{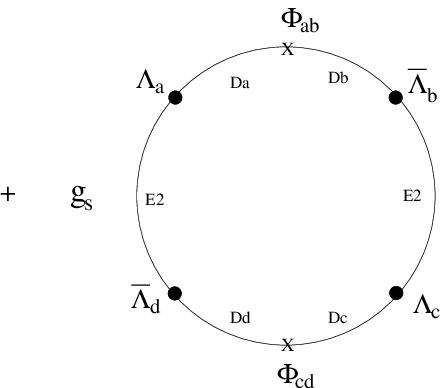} }
\caption{Leading and subleading disc diagrams for $\phi_{ab}\phi_{cd}$ coupling.
Only the leading order two-disc amplitude contributes to the superpotiential.} \label{figb}
\end{center}
\end{figure}

\noindent
As an illustration,
imagine the situation of four cycles
$\Pi_a$, $\Pi_b$, $\Pi_c$, $\Pi_d$, each intersecting an instanton $E2$
precisely once. The induced coupling for $\Phi_{ab}$ and $\Phi_{cd}$ is
given by the product of the two discs in figure \ref{figb},
whereas the one-disc amplitude  is of order $g_s$ and therefore
does not contribute to the superpotential.

\begin{figure}[h]
\begin{center}
\epsfbox{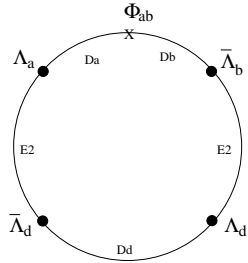}
\caption{Coupling $\phi_{ab}$ absorbing four fermionic zero modes}
        \label{figc}
\end{center}
\end{figure}

\noindent As a second example, figure \ref{figc} shows a situation where a
coupling $\phi_{ab}$ can  exist
by soaking up four fermionic zero modes, which is only possible in the presence of
both $\lambda_d$ and $\overline{\lambda}_d$ zero modes.
However, due to the $g_s^{1/2}$ factor for the fermionic zero modes, this
coupling is likewise of a higher power in $g_s$ and so does not contribute to
the superpotential.

\noindent

\begin{figure}[h]
\begin{center}
\hbox{\hskip 4cm \epsfbox{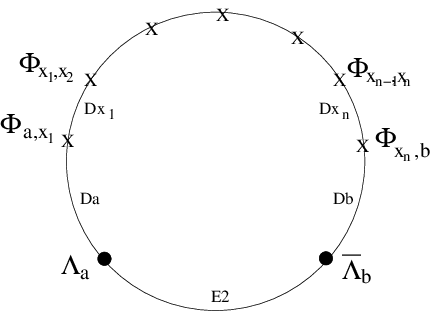}\hskip 2cm \epsfbox{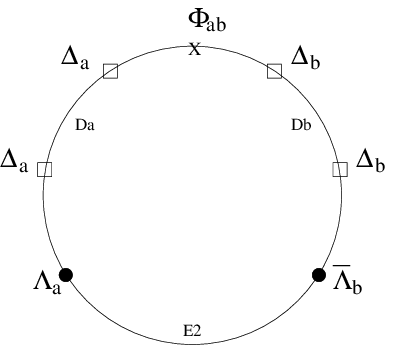} }
 \caption{Coupling $\phi_{a,x_1}\phi_{x_1, x_2}\ldots \phi_{x_n,b}$ and
        disc tadpole with D6-brane moduli insertions } \label{figcc}
\end{center}
\end{figure}

\noindent
By contrast, as depicted in  figure \ref{figcc}, it can also happen that the
zero mode structure is such that for instance a coupling
$\phi_{a,x_1}\phi_{x_1, x_2}\ldots \phi_{x_n,b}$  can be generated by soaking
up only two fermionic zero modes.

We now consider the possibility of non-rigid $D6$-branes, i.e. $b_1(\Pi_a)>0$. As anticipated, this
leads to the presence of additional $b_1(\Pi_a)$ chiral superfields in the adjoint
representation of $U({\cal N}_a)$, the open string moduli
$ \Delta_a$ \cite{McLean}. The real part of their bosonic component  parameterises
the deformations of the sLag cycle $\Pi_a$ and its imaginary part corresponds to the Wilson lines of
the $U(1)$ gauge field on a single  $D6$-brane.
Again, the superpotential has to be a holomorphic function of the full chiral superfield
$\Delta_a$, and the only possibility consistent with discrete $U(1)$ shifts in its bosonic imaginary
part is an exponential dependence \cite{Kachru:2000ih} of the form
\bea
\label{deln}
     {\rm tr} \left[ \exp\left( -{\textstyle {\Delta_a\over \alpha'}} \right)\right],
\eea
where the trace is over the CP-indices 
Such a term is due to the fact that the standard disc tadpole
diagram can  also have insertions of these moduli fields $\Delta_a$ (see figure \ref{figcc}).
In this case one has to sum over all these contributions inducing a consistent open
string moduli dependence of the superpotential couplings. As shown in
\cite{Recknagel:1998ih}, these integrable boundary deformations lead to the expected
exponential dependence.

To conclude, the only discs which can contribute to $W$ are
those with precisely two charged fermionic zero modes per disc.
It is useful to  introduce the short-hand notation
\bea
     \widehat\Phi_{a_k,b_k}[{\vec x_k}] = \Phi_{a_k,x_{k,1}}\cdot \Phi_{x_{k,1},x_{k,2}} \cdot \Phi_{x_{k,2},x_{k,3}}\cdot \ldots
             \cdot \Phi_{x_{k,n-1},x_{k,n}}     \cdot \Phi_{x_{k,n_k},b_k}
\eea
for the product of open string vertex operators from the set $\{\Phi_{a_1,b_1}, \ldots \Phi_{a_M,b_M}\}$,
where each field is allowed to  only appear once, and again consistent assignment of CP-labels is
tacitly  assumed.
We define $\widehat\Phi_{a_k,b_k}[\vec 0]=\Phi_{a_k,b_k}$.
If we have $L$ such chain-products, their lengths in an $M$-point function
have to add up to $\sum_{k=1}^L (n_k+1)=M-M_{{\rm loop}}$, where
$M_{\rm loop}$ is the number of fields attached to one-loop diagrams
(see below).

The resulting disc contribution to the instanton correlator therefore
takes the  comparably simple schematic form
\bea
&& \int d^4 x\, d^2\theta\,\,  \sum_{\rm conf.}  \, {\textstyle
  \prod_{a}  \bigl(\prod_{I=1}^{ [\Xi\cap
             \Pi_a]^+}  d\lambda_{a,I}\bigr)\,
               \bigl( \prod_{I=1}^{ [\Xi\cap
             \Pi_a]^-}  d\overline{\lambda}_{a,I}\bigr) } \, \\
   &&\phantom{aaaaaaaaaaaaa}
   \langle \widehat\Phi_{a_1,b_1}[\vec x_1]   \rangle_{\lambda_{a_1},\overline{\lambda}_{b_1}}\cdot
            \ldots \cdot  \langle \widehat\Phi_{a_L,b_L}[\vec x_L]
          \rangle_{\lambda_{a_L},\overline{\lambda}_{b_L}}
                 \, e^{-S_{E2}}. \nonumber
\eea
Here, the sum over configurations only involves the possible ways
of attaching the respective fermionic zero modes to the
matter field discs. Note that as in \cite{Cremades:2003qj,Cvetic:2003ch} each disc amplitude of boundary changing operators contains
by itself a whole sum of holomorphic disc
instantons ending on the respective branes. This introduces a dependence of the instanton amplitude
on the K\"ahler moduli $T_i=\int_{\omega^i_2} (J + i B)$ of the form
\bea
      \exp \left( {\textstyle -{1\over \alpha'} T_i } \right),
\eea
where $\omega^i_2$ denotes a basis of
the $h_{11}$ cohomological two-cycles in the CY.

\vspace{10pt}
\noindent
\underline{1-loop amplitudes} \\
\vspace{10pt}

Having discussed the disc contribution to the instanton couplings,
it only remains to clarify the role of the  1-loop diagrams.
Let us discuss the vacuum amplitudes first, where the story
is quite similar to the vacuum $(\langle 1 \rangle_{disc})$ contribution.

The relevant 1-loop diagrams are the vacuum annulus and M\"obius strip diagrams
involving at least one E2 boundary.
Here for instance an  annulus vacuum diagram $Z^A(E2,D6_a)$  is given by
\bea
     Z^A(E2,D6_a)=c\, \int_0^\infty {dt\over t}\,  {\rm Tr}_{E2,D6_a}\left( e^{-2\pi t L_0} \right),
\eea
where the trace is over all open strings stretching between the two branes.

\begin{figure}[h]
\begin{center}
\epsfbox{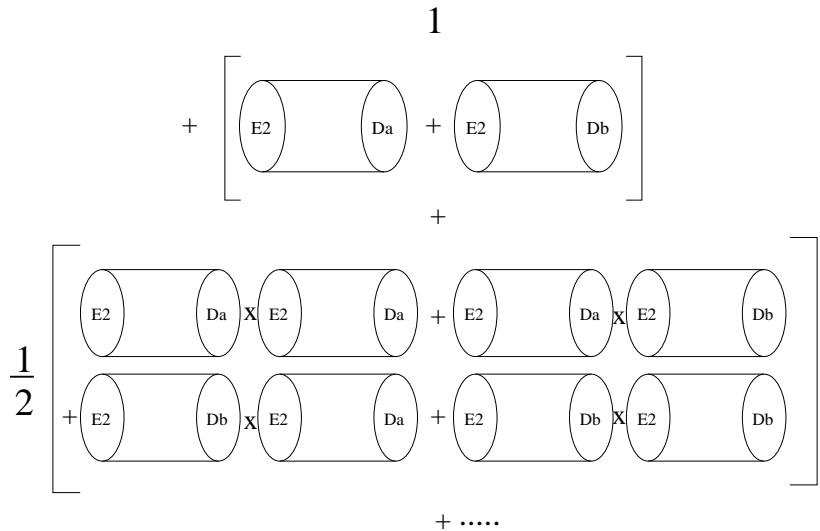} \caption{Series of 1-loop contributions for the example of only
   two D6-branes} \label{figf}
\end{center}
\end{figure}

\noindent Clearly supersymmetry implies that the amplitudes $Z^A(E2,E2)$ and $Z^A(E2,E2')$ vanish.
However, the open strings stretched between the $E2$-instanton  and a stack of $D6$-branes  preserve
only two supercharges. As we have already seen for the zero mode counting, the amplitude
is not Bose-Fermi degenerate and  we do not
expect $Z^A(E2,D6_a)$, $Z^A(E2,D6'_a)$ and the M\"obius amplitude $Z^M(E2,O6)$
to vanish.
Following the same logic as for the $(\langle 1 \rangle_{disc})^n$ terms in the instanton
amplitude, we can have all possible one-loop contributions, giving rise to
a series
\bea
\label{oneloop}
    \sum_{n=0}^\infty {1\over n!} \left( \sum_a \left[ Z^A(E2,D6_a) + Z^A(D6'_a,E2) \right]+ Z^M(E2,O6)
\right)^n=
    \exp \left( Z_0\right),
\eea
with
\bea
    && Z_0 = \sum_a \left[ Z^A(D6_a,E2)+Z^A(D6'_a,E2)\right] + Z^M(E2,O6).
\eea
This is also shown in  figure \ref{figf}.

In order to interpret this result, let us discuss the behaviour of this expression
in the limits  $t\to 0$  and $t\to \infty$.
We expect the possible $t\to 0$ divergences to
cancel by the tadpole cancellation condition for the $D6_a$ branes.
The possible $t\to \infty$ divergence is due to zero modes in the various
open string sectors.
If we have  in the NS-sector a spacetime bosonic zero mode,
 $Z_0\to +\infty$ and $\exp(Z_0)$ diverges. On the other hand, in the case of a spacetime
fermionic zero mode in the R-sector,  $Z_0\to -\infty$ and $\exp(Z_0)$ vanishes.
Moreover, if $Z_0$ vanishes identically  in a supersymmetric case, then  $\exp(Z_0)=1$.
All this is precisely the behaviour characteristic of the one-loop
Pfaffians/determinants of the kinetic operators for  the quantum fluctuation around the
E2-background.
From the saddle point approximation such terms are expected
to appear in the instanton amplitudes\footnote{In the dual heterotic model with
a world-sheet instanton wrapping a holomorphic curve $C$,
the uncharged contribution to the superpotential was argued to be \cite{Witten:1999eg}
\bea
\label{pfaff}
    W={{\rm Pfaff}(\overline\partial_{V_- } )\over  ({\rm det}'\, \overline\partial_{\cal O})^2\,
 ({\rm det}\, \overline\partial_{ {\cal O}(-1)} )^2}
   \, \exp ( -S_{\rm inst} ).
\eea
This is consistent with our picture, as
these determinants are only over the left-moving fluctuations. These involve
the vector bundle, whose role is played in our case by the D6-branes. The right-moving
heterotic fluctuations cancel due to supersymmetry and should therefore
be identified with open strings between the E2-instanton and itself. Moreover,
we propose that the $({\rm det}'\, \overline\partial_{\cal O})^2$ factor arise from the
M\"obius strip amplitude. Finally, let us point out that the connection between the Pfaffian and the exponential of a suitable trace  over the spectrum as in (\ref{oneloop}) is standard in field theory, of course.}.
Moreover, these one-loop determinants have been argued to
be exact for the contribution to the superpotential, which is confirmed
by the CFT approach.

As in \cite{Witten:1999eg}, a zero or a divergence in the one-loop amplitude (\ref{oneloop})
signals that one has integrated out a fermionic or bosonic zero mode, respectively.
Since in our amplitudes we carry out this integration over zero modes explicitly, the one-loop
term should be regularised such that these zero modes are removed from the traces.
Therefore, we propose the following intriguing relation between the spacetime
notion of 1-loop Pfaffians/determinants for quantum fluctuations around the
instanton background and the exponential of a world-sheet open string partition function
\bea
                { {\rm Pfaff'} ({\cal D}_F) \over \sqrt{ {\rm det}' ({\cal D}_B)} }=
                \exp \left( {\textstyle \sum_a \left[ {Z'}^A (E2,D6_a) +{Z'}^A (E2,D6'_a)\right] +  {Z'}^M(E2,O6)} \right).
\eea

One also has the freedom to attach the the D-brane moduli fields $\Delta_a$  to
the D6$_c$-brane boundary of the one-loop diagrams. 
After taking the sum
over all possible insertions, this leads to
the open string moduli dependence of $\exp A_0$.

\begin{figure}[h]
\begin{center}
\epsfbox{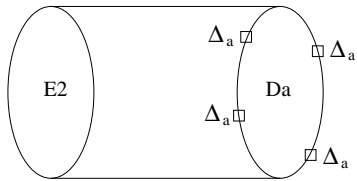} \caption{Moduli dependence of 1-loop amplitudes} \label{fige}
\end{center}
\end{figure}

\noindent
This is similar to  the case of  world-sheet instanton corrections for the
heterotic string, where
these 1-loop determinants do also depend on
the bundle and complex structure moduli \cite{Witten:1999eg}. Finally,
 we stress once more that the one-loop determinant is manifestly uncharged under the anomalous $U(1)$ gauge factors. This  direct consequence of the $g_s$-assignment for the vertex operators of the $\lambda_a$-modes is crucial for the generation of a superpotential of the form (\ref{superproduct}) - any $U(1)$ transformation of the one-loop prefactor would spoil the  gauge invariance of (\ref{superproduct}).

To summarise, the single E2-instanton contribution
to the charged matter superpotential can be determined
by evaluating the following zero mode integral over disc and 1-loop
open string CFT amplitudes
\vskip 0.3cm

\hspace*{-24pt}
\fbox{
\begin{minipage}{\textwidth}
\bea
     && \langle \Phi_{a_1,b_1}(p_1)\cdot\ldots\cdot   \Phi_{a_M,b_M}(p_M)
\rangle_{E2-{\rm inst}} = \nonumber \\
&& 
 = \int d^4 x\, d^2\theta \,\,
       \sum_{\rm conf.}\,\,  {\textstyle
  \prod_{a} \bigl(\prod_{i=1}^{ [\Xi\cap
             \Pi_a]^+}  d\lambda_a^i\bigr)\,
               \bigl( \prod_{i=1}^{ [\Xi\cap
             \Pi_a]^-}  d\overline{\lambda}_a^i\bigr) } \ \,\, \exp ({-S_{E2}}) \,
         \times \, \exp \left({Z'_0}\right) \,  \nonumber \\
&&\phantom{aaa}
\times \langle \widehat\Phi_{a_1,b_1}[\vec x_1]   \rangle_{\lambda_{a_1},\overline{\lambda}_{b_1}}\cdot
            \ldots \cdot  \langle \widehat\Phi_{a_L,b_L}[\vec x_L]
          \rangle_{\lambda_{a_L},\overline{\lambda}_{b_L}} \,\times \,
         \prod_k \langle \widehat\Phi_{c_k,c_k}[\vec x_k] \rangle^{\rm
           loop}_{A(E2,D6_{c_k})}  . \nonumber
\eea
\end{minipage}
}
\vskip 0.3cm

This formula contains from the disc the
exponential instanton action and
the combinatorics of disc tadpole diagrams with two charged
fermionic zero modes attached to each disc.
From the 1-loop diagrams, i.e. the annulus and M\"obius strip
amplitudes, there also arises  the exponential vacuum contribution,
i.e. the one-loop Pfaffian/determinant, and
the combinatorics of attaching D6-brane neutral products of vertex operators
to the boundaries of annulus diagrams.

The schematic  form of the superpotential, making explicit the moduli dependence, is
\bea
W=\sum_{E2} e^{-S_{E2}(U) }\, f\left(\exp\left( - {\textstyle {T\over \alpha'}}\right),
 {\rm tr}\left[\exp\left( - {\textstyle { \Delta\over \alpha'}}\right)\right], \Phi_{ab} \right) ,
\eea
where the exponential dependences are enforced by the fact that the imaginary
parts of the complex scalars are either axions or Wilson lines having a
shift symmetry.

In practice, to really carry out this CFT program, one needs the explicit
form of the boundary changing operators including their correct normalisation,
which are only available
in case  we really have a CFT description of the models.
We will provide more details and examples for such instanton
computations in \cite{bcw}.
For now it is a good working hypothesis that every term which can
in principle be generated, i.e. for which the zero mode counting and the combinatorics of open
strings works,  is indeed  non-vanishing.

\subsection{Comments on multi-instanton effects}
\label{multi}

In general, the non-perturbative sector may not only involve a single E2-brane, but also multi-instantons.
Of course, one expects that all these non-perturbative sectors allow a CFT description along the
lines presented so far. This includes essentially two types of generalisations.

First, the three-cycle  $\Xi$ may be wrapped by a stack of ${\cal N}_{\Xi}$ coincident E2-branes.
Clearly, in such a situation the exponential suppression (\ref{expfac}) of all resulting amplitudes
is enhanced by a factor of ${\cal N}_{\Xi}$ as compared to the single-instanton case. The fermionic zero
modes between the D6 and the E2-branes carry an additional CP-index, accounting for the fact that the
number of such zero modes is likewise multiplied by ${\cal N}_{\Xi}$. Since again each fermionic zero
mode has to be soaked up in the computation of correlators and, as we just derived, each superfield
$\Phi_{ab}$ located at the intersection of two D6-branes can absorb at most two of them for contributions
to the superpotential, these ${\cal N}_{\Xi}$-instanton sectors soon become irrelevant for the most
interesting lowest order $n$-point amplitudes as  ${\cal N}_{\Xi}$ increases.
Note also that the intersection between the cycle ${\Xi}$ and its orientifold image ${\Xi'}$ now
gives rise to fermionic zero modes in the symmetric and in the antisymmetric representation of
the $U({\cal N}_{\Xi})$ gauge group located on the stack of E2-branes.
In analogy to equation (\ref{nosyms}) we have to require that 
\bea
\label{noaantimulti}
 \Xi'\cap \Xi=0=\Pi_{O6}\cap \Xi
\eea
for the absence of these fermionic zero modes.

The most general instanton sector with possible contributions to the superpotential involves several
stacks of mutually supersymmetric E2-branes wrapping various rigid cycles ${\Xi}_i$, each with
multiplicity ${\cal N}_{\Xi_i}$. A novelty occurs since the E2-branes will generically intersect not
only the D6-branes, but also among each other. Special care is required for the string modes located at
this latter type of intersection, i.e. between all pairs of $\Xi_i$ and $\Xi_j$. This time the spectrum
is indeed Bose-Fermi degenerate due to the mutual extended supersymmetry between the E2-branes, which is a
consequence of the Dirichlet-Dirichlet boundary conditions in the non-compact four dimensions.
In particular, the massless modes now include both a pair of fermions and a complex boson and
have to be
integrated over.
Note furthermore that, due to the Bose-Fermi degeneracy between the E2-branes, the total Pfaffian simply factorises into the product of the Pfaffians resulting form the one-loop diagrams between each E2-instanton and the D6-branes.
In order to eventually extract superpotential couplings, one must only include 1PI diagrams.

\subsection{E2-instanton corrections to D-terms and the gauge kinetic function}

So far we have  focused on E2-instanton corrections to
the superpotential. From the general arguments of \cite{Kachru:2000ih,Kachru:2000an}
one also expects non-perturbative
corrections to the D-terms, which will depend on the complex structure moduli. As a consequence
there should appear E2-instanton corrections to
the  Fayet-Iliopoulos
terms for the $U(1)$s living on the D6-branes.
For all known examples perturbative and non-perturbative corrections
to the D-terms and the gauge couplings are by no means independent. In fact, the latter two quantities constitute  the argument and the absolute value
  of a complex valued central charge, see equs. (\ref{fiterm}) and (\ref{gaugecoup}).
Therefore, one likewise expects E2-instanton corrections
to the  gauge kinetic functions.
While we are not intending to compute these terms explicitly here,
we will nonetheless point our their origin and leave a more thorough
analysis for future work.

Given an intersecting D6-brane configuration, we know that
the perturbative FI terms do depend on the complex structure
moduli, which is readily understood from the corresponding dependence of the special Lagrangian condition
on these moduli. Recall from equation  (\ref{fiterm})
that the leading order (supergravity)
expression for the FI-term is proportional to the calibration condition
for the three-cycle wrapped by the E2-brane.
Assume now that the D6-branes are indeed
supersymmetric on a D-dimensional subspace $M_D\subset M_C$ of the naive
complex structure moduli space
and consider the effect of a single E2-instanton. For the generation of
F-terms we require, as described, that this instanton  wraps
a special Lagrangian  preserving the {\emph {same}} supersymmetry
as the D6-branes. In general, this instanton is only
supersymmetric on a $(D-1)$-dimensional subspace $M^{E2}_{D-1}$ of $M_C$.
If one now moves off continuously from $M^{E2}_{D-1}$ in the direction
where the D6-branes are still supersymmetric, then the instanton
no longer preserves the 4D ${\cal N}=1$ supersymmetry.
However, the number of fermionic zero modes between the D6 and the E2-brane remains unchanged
since the zero point energy in the R-sector vanishes, $E_R=0$, irrespective of the intersection angle. Secondly, and in contrast to intersecting D6-branes, the bosonic
modes still have positive mass square. This is a consequence
of the $E_{NS}\geq 0$ zero point energy in the NS-sector
and guarantees  that the D6-E2-brane configuration is still stable against
brane recombination even away fron the supersymmetric locus.

Due to the broken supersymmetry,
the instanton world-volume now accommodates four fermionic
zero modes $\theta_i$ and $\ov\theta_i$ as required for the generation of a
D-term. We conclude that this breaking of supersymmetry from the 4D point
of view
corresponds to a D-term breaking and gives rise to
the expected E2-instanton corrections to the FI terms on those  D6-branes
which are not any longer relatively supersymmetric with the $E2$-instanton.

It follows that E2-instantons wrapping sLag three-cycles which
preserve a {\emph {different}} ${\cal N}=1$ subalgebra than the D6-branes
induce instanton corrections to the D-terms.
Let us be a bit more precise and ask which further conditions the SUSY breaking
instanton sector has to meet in order to generate an FI-term.
Recall that the actual FI-term arises in the four-dimensional effective supergravity action as the term
\bea
S_{FI} = \int d^4x\, d^2 \theta\, d^2\, \ov{\theta} \ V_{U(1)},
\eea
where $V_{U(1)}$ denotes the vector superfield containing the corresponding
$U(1)$ gauge potential. What we have to compute in the CFT to probe for
such a coupling are therefore diagrams with an insertion of the vertex
operator for the auxiliary field $D_{U(1)}$ on the D6-boundary in the disc or the annulus diagram
accompanied by an even number of boundary
changing operators on the disc
corresponding to the fermionic zero modes between the
D6-brane and the SUSY-breaking instanton.

For single
instantons, this implies that the intersection between the D6 and the
 E2-brane has to be non-chiral and the E2-instanton has to be uncharged,
in agreement with the fact that the FI-term carries no $U(1)$ charge.
The combinatorics of multi-instanton contributions is richer
as now also diagrams involving several mutually non-supersymmetric
instantons contribute.

The actual computation of the diagram parallels
the story for the F-terms. The $\theta$ and $\ov \theta$ vertex operators
match with the four-dimensional spinor polarisation of the D-term component
in $V$, and all fermionic zero modes are to be integrated over. Again, the
amplitude is exponentially suppressed by the instanton action.

Similarly, inserting the vertex operators for the gauge fields
on the D6-bounda\-ries detects  the instanton corrections to the holomorphic
gauge kinetic functions. In this case, only supersymmetric configurations contribute due to the necessary appearance of precisely two $\theta$ modes on the instanton, and again the instantons have to be uncharged under the abelian gauge group.

Clearly, it would be interesting to work this out in more detail.
These corrections generalise the familiar structure known
from $\Pi$-stability \cite{Douglas:2000ah} for B-type D-branes to also include
spacetime instanton corrections. Recall that the notion of $\Pi$-stability
applies to the $g_s\to 0$ limit and adds up the world-sheet instanton
contributions to the FI-terms, which by mirror symmetry
are exact at tree level for A-type D-branes. However, the
spacetime instanton corrections are truly non-perturbative on both sides
and finding a way to compute them would be a major step forward.

\subsection{Example}
After this amusing interlude, let us go back to the superpotential and give one simple example showing  explicitly
what kind of matter couplings
can be generated  and how
the CP-factor combinatorics is taken care of.
Consider two stacks of in both cases ${\cal N}$ D6-branes wrapping the
three-cycles $\Pi_a$ and\ $\Pi_b$ with intersection number
$\Pi_a\cap \Pi_b=1$. One gets one chiral superfield $\Phi_{\overline{i},j}$
($\overline{i},j=1,\ldots,{\cal N}$)
localised at this intersection point and transforming in the
$(\overline{\cal N},{\cal N})$ representation of the $U({\cal N})\times U({\cal N})$ gauge group.

Now, imagine there exists an E2-brane wrapping a three-cycle $\Xi$
with non-vanishing intersection numbers
\bea
           [\Xi\cap \Pi_a]^+=1, \quad   [\Xi\cap \Pi_b]^-=1.
\eea
These generate in each case ${\cal N}$ zero modes $\lambda_{i}$ and
$\lambda_{\overline{j}}$. Taking into account that one chiral superfield absorbs
two fermionic zero modes, exactly one $\lambda_{i}$ and one $\lambda_{\overline{j}}$,
it is clear
that only a term $\Phi^{\cal N}$ has the potential to absorb all zero modes.
Looking more closely, and observing that interchanging the order of two
fermionic zero modes gives a relative minus sign, the coupling
introduced by this instanton has the form
\bea
             W\simeq  \det (\Phi_{\overline{i},j} )\,\, e^{-S_{E2}}
\eea
which is reminiscent of the instanton term \`a la Affleck, Dine, Seiberg \cite{Affleck:1983mk}.
Of course, here the gauge symmetry is different from the $SU(N_C)$ case of \cite{Affleck:1983mk} with $N_f=N_C-1$
non-chiral flavours.  It shows that in contrast
to perturbative string couplings, which only involve traces over Chan-Paton
factors, instantons can also generate couplings like determinants of
matter fields involving anti-symmetrisations of indices.
We will give more examples of possible couplings in section \ref{secph}.

\section{Type IIB/heterotic  dual formulations}

So far, mainly to profit from the illustrative geometric picture
we have focused on instanton effects for Type IIA
intersecting D6-brane models, but a very similar story applies
to the various dual D-brane models. In this section we would
like to  summarise aspects of instanton effects in these dual
formulations.
Application of mirror symmetry to intersecting D6-brane models leads to
Type IIB orientifolds with either O9- and O5-planes or O7- and O3-planes.

\subsection{Type IIB orientifolds with O9/O5 planes}

These  are nothing else than compactifications of the Type I string.
Here for tadpole cancellation,
one introduces D9-branes equipped with stable $U(N)$ vector bundles $V_a$ and
D5-branes wrapping effective curves of the CY threefold.
As has been summarised in \cite{Blumenhagen:2005zh} (see also \cite{Douglas:2006xy}),
the massless spectrum is given
by various cohomology classes such as $H^i({\cal X}, V_a\otimes V_b)$,
$H^i({\cal X}, \bigwedge^2 V_a)$ etc.
The Green-Schwarz mechanism works in complete analogy, now
gauging the shift symmetries of the R-R two-form $C^{(2)}$.
More concretely there are two types of axions. The first kind of axions are
given by dimensional reduction of $C^{(2)}$ on the two-cycles of the Calabi-Yau
and are called called K\"ahler axions, being  the imaginary parts of
the superfields $T$.
An additional axion arises from reducing $C^{(6)}$ on the complete CY, which leads
to the so-called universal axion sitting     in the dilaton superfield $S$.

The instantons which break these shift symmetries and
therefore the global $U(1)$s are E1-branes wrapping holomorphic two-cycles
on the CY and  E5-branes wrapping the entire CY. In general these
E5-branes can also carry stable holomorphic vector bundles so
that their Chern-Simons action involves both the K\"ahler axions and the universal
axion. The dependence of the superpotential on $T$ can therefore only be exponential.
By contrast, the complex structure $U$ and bundle moduli $B$ do
not contain any axions or Wilson-lines.

The necessary condition for E1-branes to generate a superpotential of the type described
is that they wrap rigid holomorphic curves, i.e. isolated $C=\IP^1$s.
For such an E1-brane wrapping a rigid curve $C$,
the additional charged fermionic
zero modes are counted  by the cohomology classes
$H^i (C,V_a\vert_C \otimes K_C^{1/2})$ with $i=0,1$ and
$K_{\IP^1}={\cal O}(-2)$.

For E5-instantons equipped with a gauge bundle $V_E$,
the number of extra uncharged zero modes is
given by $h^1({\cal X},V_E\otimes V_E^*)$. For higher rank bundles
this is generally non-vanishing, but for line bundles $L$ it
is given by $h^{1,0}({\cal X})=0$. The absence of zero modes
from $\Xi\cap \Xi'$ intersections yields  in this case the additional constraint
$H^i({\cal X}, {\bf S}^2 L)=0$ for $i=1,2$, with appropriate generalisations for non-abelian bundles.
The charged zero modes are counted by $H^i({\cal X},V_E\otimes
V_a)$ and  $H^i({\cal X},V_E\otimes V'_a)$ for $i=1,2$.

Again the actual instanton computation for charged matter contributions
to the superpotential boils down to the computation
of disc amplitudes for boundary changing operators.
The schematic  form of the superpotential, making explicit the moduli dependence, is
\bea
\label{superpotiib}
W=\sum_{E1} e^{ -S_{E1}(T) }\,\, f(U,B,\Phi_{ab} ) +  \sum_{E5} e^{-S_{E5}(S,T) }\, f(U,B,\Phi_{ab} ).
\eea
Notice that since the complex structure moduli do not contain any axion and the
bundle moduli $B$ no Wilson lines, the dependence on these variables does not have to be
via an exponential factor. In particular, there are no additional world-sheet instanton
corrections in an $E1$- or  $E5$-instanton sector.

\subsection{Heterotic models}
\label{sechet}

The Type I setup sketched in the previous section is S-dual to the $Spin(32)/\mbb{Z}_2$
heterotic string \cite{Witten:1984dg}, more concretely to a specific embedding of products of $U(N)$ gauge bundles into
$Spin(32)/\mbb{Z}_2$\cite{Blumenhagen:2005pm,Blumenhagen:2005zg}. A very similar story is likewise
expected in the context of $E_8 \times E_8$ heterotic compactifications with anomalous $U(1)$
factors as they generically appear upon embedding line or $U(N)$ vector bundles into $E_8 \times E_8$
\cite{Distler:1987ee,Blumenhagen:2005ga}. In the heterotic picture one has world-sheet instantons
and Euclidean 5-brane instantons with the general moduli dependence being as in
(\ref{superpotiib}).
The concrete computation
is different, of course, since there are no D-branes and therefore no disc diagrams
to be computed. In particular for the heterotic 5-brane instantons,
it is not clear how a concrete computation can be performed, as
the vertex operators are not known.

World-sheet instantons have been discussed in a couple of papers
\cite{Dine:1986zy,Dine:1987bq,Distler:1987ee,Witten:1999eg,Buchbinder:2002ic}.
For certain kinds
of bundles, namely those which are described in the (half) linear sigma model
approach, the authors of \cite{Silverstein:1995re,Basu:2003bq,Beasley:2003fx,Beasley:2005iu} have been able to prove that the various contributions to the superpotential
exactly cancel.
It remains to be seen whether such strong statements can also be made
for certain classes of for instance intersecting D6-brane models.

\subsection{Type IIB orientifolds with O7/O3-planes}

These orientifolds with D7- and D3-branes have extensively been studied
recently, in particular as they allow a very clean introduction
of flux compactifications with frozen complex structure and dilaton
moduli. Most importantly,
 E3-brane instantons can also freeze the K\"ahler moduli \cite{Witten:1996bn}, as applied e.g. in the KKLT scenario \cite{Kachru:2003aw}.
Before we come back to this latter point, let us first discuss
the general framework of instanton effects.

Similarly to the Type I case, the D7-branes wrap holomorphic divisors, i.e.
four-cycles $\Gamma_a$ in ${\cal X}$ and also carry non-trivial bundles $V_a$.
Here the dimensionally reduced
R-R four-form $C_4$ provides the four-dimensional axions to be
gauged under the $U(1)_a$ gauge groups living on the stacks of
D7-branes.

Therefore, the role of the instantons  generating the  charged matter couplings
is played by E3-branes wrapping
holomorphic four-cycles $\Gamma_E$ on the CY threefold.
Again in general these E3-branes can be equipped with holomorphic
stable vector bundles $V_E$.
To avoid neutral extra zero modes, the four-cycles $\Gamma_E$
should be rigid, i.e. $h^0(\Gamma_E, N)=0$, which means
$h^{(2,0)}(\Gamma_E)=0$. In addition there can  be bundle moduli,
which for line bundles $V_E=L$ are absent if  $h^{(1,0)}(\Gamma_E)=0$.
There can also be 4D space time filling D3-branes and E(-1)-instantons,
but these lead to non-chiral matter and only give rise to massless
modes if the branes are  localised at the same point in the transversal space.

Additional charged zero modes are given by non-trivial intersections of
the E3-branes with the D7-branes, which can then be compensated
by extra matter fields in the instanton correlator.
It has already been anticipated in \cite{Ganor:1996pe} that
if a D3-brane comes together with an E3-brane, there appears
a pair $\lambda_{3},\overline{\lambda}_3$ of fermionic zero modes.
As has been pointed out in the literature \cite{Achucarro:2006zf}, for the generalised
KKLT \cite{Kachru:2003aw} scenario proposed in \cite{Burgess:2003ic}
the extra fermionic zero modes pose a problem with the
up-lifting mechanism by magnetised D7-branes.
Since in this simplified model the CY  only has one four-cycle, both
the E3-instanton as well as the uplifting D7-brane are wrapping
the same four-cycle, leading necessarily to extra
charged zero modes on the instanton world-volume.
Therefore, in this model the E3-brane does
not generate a  purely exponential term in the superpotential;
the induced non-perturbative terms rather involve appropriate extra charged matter
fields (see \cite{dieter} for a similar discussion about a non-perturbative
superpotential via gaugino condensation on D7-branes).

To conclude, not only the background fluxes have the potential to change the zero
mode structure on the E3-instanton
\cite{Bergshoeff:2005yp,Lust:2005bd,Lust:2005cu}, but likewise possibly present
D7- and D3-branes.

\section{Phenomenological implications}
\label{secph}

Building upon our analysis in section \ref{E2_gen},
we now point out some immediate phenomenological
consequences of  these instanton generated terms in
the superpotential. The first one concerning a new destabilisation mechanism is rather worry-some.
Our second observation on the generation of perturbatively absent couplings is relevant  for proton decay, $\mu$-terms in the Higgs sector and  right-handed neutrino masses
in MSSM-like intersecting D6-brane models.  In the latter two cases, instantons provide a stringy mechanism to
explain  hierarchies.

\subsection{Vacuum destabilisation}
\label{secstable}

As explained,
for all known classes of ${\cal N}=1$ supersymmetric four-dimensional
string vacua
spacetime instanton effects have the potential to not only generate new
couplings in the theory which break the global $U(1)$ symmetries, but
they can also generate a superpotential leading (without any
other stabilising effects) to a runaway behaviour for some of the
scalar modes. That means that such  terms generate a tadpole for
an auxiliary $F$-field and therefore break supersymmetry.

Clearly, those terms can be generated if the E2-instanton does not carry
any additional uncharged and charged fermionic zero modes, i.e. if it is rigid and
does not intersect any of the D6-branes. This is the analogy
of the condition for the heterotic string of Distler/Greene \cite{Distler:1987ee} that there be no
additional left-moving fermionic zero modes on the world-sheet
instanton wrapping an isolated curve in the CY.
However, we would like to investigate the question whether instantons
which carry additional zero modes cannot destabilise the vacuum as well.

First, if there are the right number of charged fermionic zero modes with
just two  $U(1)$ stacks of D6-branes, a tadpole of the form
\bea
            W= \Phi_{a,b}\,\, e^{-S_{E2}}
\eea
for a charged matter field can be generated. Expanding around $\Phi_{a,b}=0$ we see that
even though $W$ vanishes there, $\partial W$ does not and therefore supersymmetry is broken.

\begin{figure}[h]
\begin{center}
\epsfbox{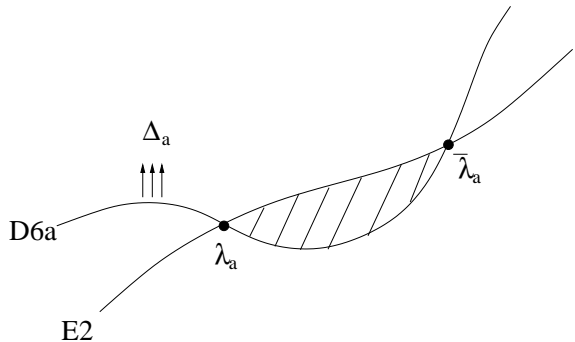}
\caption{Disc instability}
        \label{figtad}
\end{center}
\end{figure}

Second, one can imagine the situation depicted in figure \ref{figtad},
where a pair of charged fermionic zero modes is soaked up
by for instance an open string modulus $\Delta_a$ of a D6-brane.
All these terms lead to an instability for the D6-brane modulus such that the D6-branes gets pulled away from the E2-brane
in order to minimise the energy of the disc spanned between them.
Here again $\partial_{\Delta_a} W$ does not vanish. 
The effect of the resulting non-perturbative F-term for the open string moduli is expected to be comparable to the one analysed in \cite{Baumann:2006th} on the mirror symmetric Type IIB side.

\noindent
From these results we draw the conclusion that the presence  of additional zero modes
alone is not a sufficient condition for the stability of the background.
Additional zero modes can be soaked up
in a way that still leads to a destabilisation of the model.
This picture is consistent with the explicitly computed
heterotic superpotential in \cite{Buchbinder:2002ic}. There the one-loop determinants
are indeed (vector bundle) moduli dependent and their zeros are precisely
at the locations where extra massless modes appear. However, due to the moduli dependence,
$\partial W$ could nevertheless be non-vanishing, and it is precisely the derivatives of the superpotential which our correlation functions probe for.

In general one expects such destabilising terms  not to be completely absent and
it is far from obvious that summing up all these terms leads
to unexpected cancellations like for world-sheet instantons
in the context of  certain heterotic vacua
\cite{Silverstein:1995re,Beasley:2003fx,Beasley:2005iu}.
At least, for toroidal models or the $\mbb{Z}_2\times \mbb{Z}_2$
orbifold in \cite{Cvetic:2001nr,Cvetic:2001tj} one seems
to be on the safe side as all  sLag three-cycles known to us are non-rigid.

\subsection{Instanton generated matter couplings}

To be a little bit more concrete, let us consider MSSM like models
of intersecting D6-branes.
It is known that one
economic way to realize such a model is to start with four stacks of
D-branes with initial gauge symmetry \cite{Cremades:2002va}
\bea
U(3)\times Sp(2)\times U(1)\times U(1).
\eea
The Standard Model matter fields can be realized by bifundamental
fermions as shown in Table \ref{tferm}. Note that in this case there
are no chiral fermions transforming in the symmetric or
antisymmetric representations of the unitary gauge symmetries.
We also assume the existence of one vector-like pair of weak Higgs doublet
$H^\pm$ from the $(bc)$ sector.
In such a model one has three $U(1)$ gauge symmetries, of which
the  linear combination
\bea
\label{hyperc}
    U(1)_Y={1\over 3}\, U(1)_a - U(1)_c + U(1)_d
\eea
is required to stay massless, i.e. it is in the kernel of the
matrix ${\cal Q}$ in (\ref{gaugetrafob}).

\begin{table}[h]
\centering \vspace{3mm}
\begin{tabular}{|c|c|c|c|}
\hline
Intersection & Matter & Rep. & $Y$  \\
\hline\hline
$(a,b)$ & $Q_L$ & $3\times (3,2)_{(1,0,0)}$ \hfill & $1/3$ \\
$(a,c)$ & $(U_R)^c$ & $3\times (\overline 3,1)_{(-1,1,0)}$ & $-4/3$ \\
$(a',c)$ & $(D_R)^c$ & $3\times (\overline 3,1)_{(-1,-1,0)}$ & $2/3$ \\
$(b,d)$ & $L_L$ & $3\times (1,2)_{(0,0,-1)}$ & $-1$ \\
$(c,d)$ & $(E_R)^c$ & $3\times (1,1)_{(0,-1,1)}$ & $2$ \\
$(c',d)$ & $(N_R)^c$ & $3\times (1,1)_{(0,1,1)}$ & $0$ \\
\hline
\end{tabular}
\caption{Chiral massless spectrum MSSM four stack model.
$(.)^c$ denotes the charge conjugated field.
\label{tferm}}
\end{table}

Since we assume that we get just the MSSM spectrum from such a brane
configuration, we require furthermore that all brane volumes are equal.
This means that our vacuum really exhibits gauge coupling unification at the GUT
scale, which in Type IIA we can identify with the string scale
by tuning $g_s$ accordingly. Moreover, we can also choose
the Kaluza-Klein scale to be the string scale.
Suppose now that we {\it  {can}} find a completely perfect
intersecting D-brane model.

Since $U(1)_Y$ is required to stay massless, the net hypercharge
of all fermionic zero modes is guaranteed to vanish. This is simply
a consequence of the fact that the linear combination (\ref{hyperc})
is in the kernel of the
matrix ${\cal Q}$ in (\ref{gaugetrafob}).
Therefore it is clear that all instanton-generated chiral matter
couplings in the superpotential respect the gauge symmetries
of the Standard Model. Only the global symmetries like baryon
number $Q_B=Q_a$ or lepton number $Q_L=Q_d$ can be broken.
For couplings which are not forbidden in perturbation theory, i.e.
those respecting all global $U(1)$ symmetries, the non-perturbative
effects are expected to give merely small corrections to the
perturbative values. However, for couplings
generated only non-perturbatively, these are the leading order effects
and might have interesting phenomenological consequences.

For instance,  if the zero modes match, a dangerous tadpole can only be
generated for the right-handed neutrino.
More generally, all neutrino couplings
\bea
           W\simeq \left[(N_R)^c\right]^n\, e^{-S_{E2}},\quad n=0,1,2,\ldots
\eea
can in principle be generated, including also a Majorana mass term.
We will discuss such terms in more detail in section \ref{secneut}.

Another important coupling which is  absent in perturbation theory is a
$\mu$-term $\mu H^+ \, H^-$ coupling, which   involves the  Standard Model
Higgs pairs $H^{\pm}$ that originate  in an $N=1$ chiral sector, e.g.  as in
systematic constructions of three-family supersymmetric Standard Models
  in \cite{Cvetic:2004ui}. Here, suitable instantons can
likewise induce exponentially suppressed couplings and therefore explain the
hierarchically small $\mu$-term for appropriate choices of the three-cycle
volumes.

We would also like to point out that, however, in other situations the
restrictive fermionic zero-mode structure prohibits the generation of
perturbatively forbidden, but phenomenologically desirable terms such as the
tri-linear Yukawa couplings ${\bf 10}\,  {\bf 10} \, {\bf 5}_H$ in certain
intersecting brane Grand Unified  $SU(5)$ models.

Since exemplifying these two latter points would require some discussion of the explicit realisation of the involved matter fields in the various setups, we prefer to return to our prototype MSSM spectrum in table \ref{tferm} and analyse the possibility of generating non-perturbative potentially dangerous dimension-four proton decay couplings and Majorana masses in more detail. The discussion of the other couplings in slightly different settings will be exactly along these lines.

\subsection{Proton decay}

In this paper we do not intend to give a complete discussion about
proton decay, but only  discuss the dimension-four
couplings
\bea
       W_4=  \lambda\, [Q_L\, (D_R)^c\, L_L] +
       \lambda'\, [ (U_R)^c\, (D_R)^c\, (D_R)^c ] + \lambda''\, [L_L\, L_L\, (E_R)^c ],
\eea
which are usually absent in the MSSM due to R-parity.
Taking into account that each matter field soaks up two charged fermionic
zero modes and that the number of these zero modes scales with the
number of branes  ${\cal N}_a$ in a stack, one can see that the first
coupling  $[Q_L\, (D_R)^c\, L_L]$  cannot be generated by E2-instantons.
The second coupling for instance can be generated by an E2-instanton wrapping
a three-cycle $\Xi$
with the non-vanishing intersections
\bea
   [\Xi\cap \Pi_a]^+=1, \  [\Xi\cap \Pi_c]^-=1,\
      [\Xi\cap \Pi'_c]^-=2
\eea
giving rise to six fermionic zero modes.
Of course, the last two couplings
are suppressed by the instanton action
\bea
    e^{-{2\pi\over \ell_s^3\, g_s} {\rm Vol}_{E2} }.
\eea
However, proton decay due to the dimension four couplings above always
involves a contribution from the first coupling. In fact, the phenomenological
bounds from these couplings always involve the product $\lambda\, \lambda'$
\cite{Nath:2006ut} and therefore, even after taking the E2-instanton corrections into
account, the proton is stable in the MSSM intersecting D-brane model.
Other phenomenological bounds for  baryon and lepton  number violating
processes induced by the dimension four couplings $\lambda'$ and $\lambda''$
require a rather mild exponential suppression \cite{Allanach:2003eb}, which
should be naturally generated by appropriate instanton actions.

\subsection{Neutrino masses}
\label{secneut}

The other sector where non-perturbative effects are important is the neutrino
sector. As we have seen, these MSSM-like intersecting D6-brane models
automatically contain right-handed neutrinos so that one can contemplate
implementing the seesaw mechanism in this setup \footnote{For an alternative
proposal for small left-handed neutrino  Majorana-like masses within  $SU(5)$
Grand Unified  models with intersecting D6-branes, see
\cite{CveticLangacker}.}.

 First the Dirac mass terms are generated
perturbatively via the weak Higgs effect \bea
          W_H= H^+\, L_L\, (N_R)^c,
\eea
so that as usual one expects these mass terms to be of the order of the
quark/lepton masses.
Since $L_L$ carries non-vanishing $U(1)_Y$ charge, there are no
E2-instanton contributions to the Majorana mass of the left-handed
neutrinos. There might be higher-dimensional couplings but
these will be further suppressed with $M_s$.

On the contrary, as we discussed, E2-instanton effects
can generate Majorana type masses for the right-handed neutrinos
\bea
            W_m= M_m\, (N_R)^c\, (N_R)^c
\eea
with the mass scale
\bea
           M_m=x\, M_s\, e^{-{2\pi\over \ell_s^3\, g_s} {\rm Vol}_{E2} },
\eea
where $x$ summarises the contributions from the K\"ahler potential
(normalisation of kinetic terms) and the one-loop determinant.
One does not expect these contributions to be exponentially
suppressed or enhanced so we assume that $x$ is of order $O(1)$.
Such a mass-term satisfies all selection rules if the intersection
numbers of the rigid E2-instanton with the MSSM branes are
\bea
   &&\Xi\cap \Pi_a=\Xi\cap \Pi'_a=\Xi\cap \Pi_b=\Xi\cap \Pi'_b=
    \Xi\cap \Pi_c=\Xi\cap \Pi'_d=0 \nonumber\\
   &&[\Xi\cap \Pi'_c]^+=2,\,\, [\Xi\cap \Pi'_c]^-=0,\,\, [\Xi\cap \Pi_d]^- =2,\,\,
   [ \Xi\cap \Pi_d]^+ =0.
\eea

Note that even though instanton effects are genuinely small, they
are small compared to the natural scale in the problem,
which here is the stringy  mass scale, i.e. $M_s$
\footnote{Our whole argument here relies on the assumption that the
overall mass scale of this term is $M_s$. However, we  only   see one other
stringy mass scale in the problem and that is the KK scale, which
was also assumed to be of order $M_s$. Due to the purely stringy origin
of this term, the weak scale is not a natural scale to appear here,
as it only appears after supersymmetry breaking.}.

Thus, we have now all ingredients together for the seesaw mechanism to work.
The exponential suppression from the instanton action allows one,
without any fine-tuning,
 to explain a possible  hierarchy between the string or
GUT scale and the Majorana mass scale, which phenomenologically lies in the
range $10^{11}{\rm GeV}<M_m<10^{15}{\rm GeV}$.

To get a feeling of the scales involved let us assume
that one really found a four-stack intersecting D-brane model,
which  gives just the MSSM and features gauge coupling unification
at the usual GUT scale of $M_{\rm GUT}=2\cdot 10^{16}$ GeV, i.e.
all the volumes of the three-cycle wrapped by the MSSM branes
are equal.
Then the gauge coupling at the GUT scale is given by
\bea
   {1\over \alpha_{\rm GUT}}= {1\over \ell_s^3\, g_s} {\rm Vol}_{D6}
\eea
which implies the following relation for the neutrino masses
\bea
      M_m=x\, M_s\,  e^{-{2\pi\over \alpha_{\rm GUT}} { {\rm Vol}_{E2}\over
              {\rm Vol}_{D6} }}.
\eea
Assuming $x=1$ and   isotropic three-cycles, the phenomenological bounds
on the Majorana mass scale implies the following
moderate bounds for the length scales  of the three-cycles wrapped by the MSSM
branes and the instanton
\bea
        0.4\cdot R_{D6}> R_{E2} > 0.27\cdot R_{D6}.
\eea

\section{Conclusions}

For essentially all   known phenomenologically interesting classes of supersymmetric four-di\-men\-sional
string compactifications we have investigated
the effects  of non-perturbative spacetime instantons on the superpotential couplings for the charged
matter superfields. We pointed out that the formal global $U(1)$ charges
carried by the exponential instanton factor are due to additional fermionic
zero modes on the world-volume of the instanton which arise from
its intersection with the other space time filling D-branes in the model.
We provided an outline of the concrete computation of such instanton amplitudes
and argued that eventually it boils down to the computation
of open string disc diagrams for boundary changing operators. These are multiplied
by the one-loop vacuum diagram, where only those open string sectors
preserving exactly two  supercharges contribute.
We briefly  pointed out some further applications of these methods including the computation of multi-instanton contributions
and the important spacetime instanton effects modifying the D-terms
and the gauge kinetic functions.
Whereas we argued that these latter non-perturbative corrections to the supersymmetry conditions are not expected to destabilize the vacuum drastically, it is nonetheless important to analyse their effects more quantitatively. In particular, it remains for future work to investigate possible consequences on the statistical distribution of supersymmetric brane vacua in the type IIA geometric regime \cite{Blumenhagen:2004xx,Gmeiner:2005vz,Douglas:2006xy} or at the small radius Gepner point \cite{Dijkstra:2004cc,Anastasopoulos:2006da}.

We explained how E2-instanton effects in Type IIA orientifolds are mapped via T-duality
to essentially E1/E5- and E3-instanton effects in Type IIB orientifolds.
These latter effects  have been important for recent discussions on moduli stabilisation
and we pointed out that our general investigation forbids the D-term uplifting mechanism
in the KKLT scenario.

Finally, we applied this framework to MSSM like intersecting D-brane models
and argued that besides destabilising terms also new couplings violating
the global $U(1)$ symmetries can be generated. Such couplings can
induce  Majorana mass terms for the right-handed neutrinos, thus
providing the essential ingredient for the seesaw mechanism.

In this paper we focused mainly on single instantons which do not
have any additional bosonic zero modes since they are wrapping rigid cycles
in the internal CY manifold. Instantons carrying  additional zero modes
can result in more general superpotential couplings presumably
mixing charged matter and closed string moduli fields.
Moreover, here we have only presented  the general description of how such
instanton computations can be carried out in principle. It would be
interesting to apply these methods to concrete exactly solvable
conformal field theory models like toroidal orbifold and Gepner models.

It is important to see  whether the heterotic instanton sum rules of \cite{Beasley:2003fx}
can be generalised to D-brane models, thus ensuring that they
are not destabilised by instantons. Of course combining these
non-perturbative superpotential terms with flux generated terms
can also lead to   a stabilisation of  (all) moduli in certain string models.

The F- and D-term instanton corrections
discussed in this paper  point towards  a finite $g_s$ generalisation of
${\cal N}=1$ mirror symmetry including not only world-sheet but also spacetime
instanton corrections. Such a mathematical structure, involving
a spacetime
non-perturbative extension of Fukaya/derived categories,  seems
to underly  a
rigorous approach to the string vacuum problem and eventually
the discipline of what should really be  called string phenomenology.

\vskip 1cm
 {\noindent  {\Large \bf Acknowledgements}}
 \vskip 0.5cm
We gratefully acknowledge discussions with Ron Donagi, Michael Haack, Shamit Kachru, Paul Langacker, Tao Liu, Dieter L\"ust, Tony Pantev, Michael Pl\"umacher, Robert Richter, Maximilian Schmidt-Sommerfeld and
Marco Zagermann.
R.B. and M.C. would like to thank the KITP at the University of California
at Santa Barbara  for hospitality.
This research was supported in part by the National Science Foundation under
Grant No. PHY99-07949, the
Department of Energy Grant
DOE-EY-76-02-3071, the Fay R. and Eugene L. Langberg
Endowed Chair and the National Science Foundation  Grant
INT02-03585 (M.C. and T.W.).

\clearpage
\nocite{*}
\bibliography{rev}

\providecommand{\href}[2]{#2}\begingroup\raggedright\begin{thebibliography}{10}

\bibitem{Dine:1986zy}
M.~Dine, N.~Seiberg, X.~G. Wen, and E.~Witten, ``Nonperturbative effects on the
  string world sheet,'' {\em Nucl. Phys.} {\bf B278} (1986)
769.

\bibitem{Dine:1987bq}
M.~Dine, N.~Seiberg, X.~G. Wen, and E.~Witten, ``Nonperturbative effects on the
  string world sheet. 2,'' {\em Nucl. Phys.} {\bf B289} (1987)
319.

\bibitem{Distler:1987ee}
J.~Distler and B.~R. Greene, ``Aspects of (2,0) compactifications,'' {\em Nucl.
  Phys.} {\bf B304} (1988)
1.

\bibitem{Witten:1999eg}
E.~Witten, ``World-sheet corrections via D-instantons,'' {\em JHEP} {\bf 02}
  (2000) 030,
\href{http://www.arXiv.org/abs/hep-th/9907041}{{\tt hep-th/9907041}}.

\bibitem{Buchbinder:2002ic}
E.~I. Buchbinder, R.~Donagi, and B.~A. Ovrut, ``Superpotentials for vector
  bundle moduli,'' {\em Nucl. Phys.} {\bf B653} (2003) 400--420,
\href{http://www.arXiv.org/abs/hep-th/0205190}{{\tt hep-th/0205190}}.

\bibitem{Buchbinder:2002pr}
E.~I. Buchbinder, R.~Donagi, and B.~A. Ovrut, ``Vector bundle moduli
  superpotentials in heterotic superstrings and M-theory,'' {\em JHEP} {\bf 07}
  (2002) 066,
\href{http://www.arXiv.org/abs/hep-th/0206203}{{\tt hep-th/0206203}}.

\bibitem{Kachru:2000ih}
S.~Kachru, S.~Katz, A.~E. Lawrence, and J.~McGreevy, ``Open string instantons
  and superpotentials,'' {\em Phys. Rev.} {\bf D62} (2000) 026001,
\href{http://www.arXiv.org/abs/hep-th/9912151}{{\tt hep-th/9912151}}.

\bibitem{Kachru:2000an}
S.~Kachru, S.~Katz, A.~E. Lawrence, and J.~McGreevy, ``Mirror symmetry for open
  strings,'' {\em Phys. Rev.} {\bf D62} (2000) 126005,
\href{http://www.arXiv.org/abs/hep-th/0006047}{{\tt hep-th/0006047}}.

\bibitem{Aganagic:2000gs}
M.~Aganagic and C.~Vafa, ``Mirror symmetry, D-branes and counting holomorphic
  discs,''
\href{http://www.arXiv.org/abs/hep-th/0012041}{{\tt hep-th/0012041}}.

\bibitem{Becker:1995kb}
K.~Becker, M.~Becker, and A.~Strominger, ``Five-branes, membranes and
  nonperturbative string theory,'' {\em Nucl. Phys.} {\bf B456} (1995)
  130--152,
\href{http://www.arXiv.org/abs/hep-th/9507158}{{\tt hep-th/9507158}}.

\bibitem{Harvey:1999as}
J.~A. Harvey and G.~W. Moore, ``Superpotentials and membrane instantons,''
\href{http://www.arXiv.org/abs/hep-th/9907026}{{\tt hep-th/9907026}}.

\bibitem{Belani:2006wx}
K.~Belani, P.~Kaura, and A.~Misra, ``Supersymmetry of noncompact MQCD-like
  membrane instantons and heat kernel asymptotics,''
\href{http://www.arXiv.org/abs/hep-th/0603063}{{\tt hep-th/0603063}}.

\bibitem{Green:1997tv}
M.~B. Green and M.~Gutperle, ``Effects of D-instantons,'' {\em Nucl. Phys.}
  {\bf B498} (1997) 195--227,
\href{http://www.arXiv.org/abs/hep-th/9701093}{{\tt hep-th/9701093}}.

\bibitem{Gutperle:1997iy}
M.~Gutperle, ``Aspects of D-instantons,''
\href{http://www.arXiv.org/abs/hep-th/9712156}{{\tt hep-th/9712156}}.

\bibitem{Billo:2002hm}
M.~Billo {\em et al.}, ``Classical gauge instantons from open strings,'' {\em
  JHEP} {\bf 02} (2003) 045,
\href{http://www.arXiv.org/abs/hep-th/0211250}{{\tt hep-th/0211250}}.

\bibitem{Kashani-Poor:2005si}
A.-K. Kashani-Poor and A.~Tomasiello, ``A stringy test of flux-induced isometry
  gauging,'' {\em Nucl. Phys.} {\bf B728} (2005) 135--147,
\href{http://www.arXiv.org/abs/hep-th/0505208}{{\tt hep-th/0505208}}.

\bibitem{Witten:1996bn}
E.~Witten, ``Non-Perturbative Superpotentials In String Theory,'' {\em Nucl.
  Phys.} {\bf B474} (1996) 343--360,
\href{http://www.arXiv.org/abs/hep-th/9604030}{{\tt hep-th/9604030}}.

\bibitem{Blumenhagen:2005mu}
R.~Blumenhagen, M.~Cveti{\v c}, P.~Langacker, and G.~Shiu, ``Toward realistic
  intersecting D-brane models,''
\href{http://www.arXiv.org/abs/hep-th/0502005}{{\tt hep-th/0502005}}.

\bibitem{bcw}
R.~Blumenhagen, M.~Cveti{\v c}, and T.~Weigand, ``in preparation,''.

\bibitem{Shamit}
B.~Florea, S.~Kachru, J.~McGreevy, and N.~Saulina, ``Stringy Instantons and
  Quiver Gauge Theories,''
\href{http://www.arXiv.org/abs/hep-th/0610003}{{\tt hep-th/0610003}}.

\bibitem{dieter}
M.~Haack, D.~Krefl, D.~Lust, A.~Van~Proeyen, and M.~Zagermann, ``Gaugino
  Condensates and D-terms from D7-branes,''
\href{http://www.arXiv.org/abs/hep-th/0609211}{{\tt hep-th/0609211}}.

\bibitem{McGreevy}
J.~McGreevy, ``On the Capture of Runaway Quivers,'' {\em Talk given at Strings
  2006, Beijing}
  \href{http://www.arXiv.org/abs/http://strings06.itp.ac.cn/talk-files/mcgreev%
y.pdf}{{\tt http://strings06.itp.ac.cn/talk-files/mcgreevy.pdf}}.

\bibitem{Blumenhagen:2002wn}
R.~Blumenhagen, V.~Braun, B.~K{\"o}rs, and D.~L{\"u}st, ``Orientifolds of K3
  and Calabi-Yau manifolds with intersecting D-branes,'' {\em JHEP} {\bf 07}
  (2002) 026,
\href{http://www.arXiv.org/abs/hep-th/0206038}{{\tt hep-th/0206038}}.

\bibitem{Aldazabal:2000dg}
G.~Aldazabal, S.~Franco, L.~E. Ibanez, R.~Rabadan, and A.~M. Uranga, ``D = 4
  chiral string compactifications from intersecting branes,'' {\em J. Math.
  Phys.} {\bf 42} (2001) 3103--3126,
\href{http://www.arXiv.org/abs/hep-th/0011073}{{\tt hep-th/0011073}}.

\bibitem{Klebanov:2003my}
I.~R. Klebanov and E.~Witten, ``Proton decay in intersecting D-brane models,''
  {\em Nucl. Phys.} {\bf B664} (2003) 3--20,
\href{http://www.arXiv.org/abs/hep-th/0304079}{{\tt hep-th/0304079}}.

\bibitem{Blumenhagen:2003jy}
R.~Blumenhagen, D.~L{\"u}st, and S.~Stieberger, ``Gauge unification in
  supersymmetric intersecting brane worlds,'' {\em JHEP} {\bf 07} (2003) 036,
\href{http://www.arXiv.org/abs/hep-th/0305146}{{\tt hep-th/0305146}}.

\bibitem{Font:2006na}
A.~Font, L.~E. Ibanez, and F.~Marchesano, ``Coisotropic D8-branes and
  model-building,''
\href{http://www.arXiv.org/abs/hep-th/0607219}{{\tt hep-th/0607219}}.

\bibitem{Polchinski:1994fq}
J.~Polchinski, ``Combinatorics of boundaries in string theory,'' {\em Phys.
  Rev.} {\bf D50} (1994) 6041--6045,
\href{http://www.arXiv.org/abs/hep-th/9407031}{{\tt hep-th/9407031}}.

\bibitem{Beasley:2005iu}
C.~Beasley and E.~Witten, ``New instanton effects in string theory,'' {\em
  JHEP} {\bf 02} (2006) 060,
\href{http://www.arXiv.org/abs/hep-th/0512039}{{\tt hep-th/0512039}}.

\bibitem{Blumenhagen:2005tn}
R.~Blumenhagen, M.~Cveti{\v c}, F.~Marchesano, and G.~Shiu, ``Chiral D-brane
  models with frozen open string moduli,'' {\em JHEP} {\bf 03} (2005) 050,
\href{http://www.arXiv.org/abs/hep-th/0502095}{{\tt hep-th/0502095}}.

\bibitem{Blumenhagen:2006ab}
R.~Blumenhagen and E.~Plauschinn, ``Intersecting D-branes on shift Z(2) x Z(2)
  orientifolds,'' {\em JHEP} {\bf 08} (2006) 031,
\href{http://www.arXiv.org/abs/hep-th/0604033}{{\tt hep-th/0604033}}.

\bibitem{Berkooz:1996km}
M.~Berkooz, M.~R. Douglas, and R.~G. Leigh, ``Branes intersecting at angles,''
  {\em Nucl. Phys.} {\bf B480} (1996) 265--278,
\href{http://www.arXiv.org/abs/hep-th/9606139}{{\tt hep-th/9606139}}.

\bibitem{Ganor:1996pe}
O.~J. Ganor, ``A note on zeroes of superpotentials in F-theory,'' {\em Nucl.
  Phys.} {\bf B499} (1997) 55--66,
\href{http://www.arXiv.org/abs/hep-th/9612077}{{\tt hep-th/9612077}}.

\bibitem{Cvetic:2003ch}
M.~Cveti{\v c} and I.~Papadimitriou, ``Conformal field theory couplings for
  intersecting D-branes on orientifolds,'' {\em Phys. Rev.} {\bf D68} (2003)
  046001,
\href{http://www.arXiv.org/abs/hep-th/0303083}{{\tt hep-th/0303083}}.

\bibitem{Cvetic:2006iz}
M.~Cveti{\v c} and R.~Richter, ``Proton decay via dimension-six operators in
  intersecting D6-brane models,''
\href{http://www.arXiv.org/abs/hep-th/0606001}{{\tt hep-th/0606001}}.

\bibitem{Giddings:2005ff}
S.~B. Giddings and A.~Maharana, ``Dynamics of warped compactifications and the
  shape of the warped landscape,'' {\em Phys. Rev.} {\bf D73} (2006) 126003,
\href{http://www.arXiv.org/abs/hep-th/0507158}{{\tt hep-th/0507158}}.

\bibitem{Brunner:1999jq}
I.~Brunner, M.~R. Douglas, A.~E. Lawrence, and C.~Romelsberger, ``D-branes on
  the quintic,'' {\em JHEP} {\bf 08} (2000) 015,
\href{http://www.arXiv.org/abs/hep-th/9906200}{{\tt hep-th/9906200}}.

\bibitem{McLean}
R.~Mc~Lean, ``Deformations of Calibrated Submanifolds,'' {\em Comm.Anal.Geom.}
  {\bf 6} (1998) 705.

\bibitem{Recknagel:1998ih}
A.~Recknagel and V.~Schomerus, ``Boundary deformation theory and moduli spaces
  of D-branes,'' {\em Nucl. Phys.} {\bf B545} (1999) 233--282,
\href{http://www.arXiv.org/abs/hep-th/9811237}{{\tt hep-th/9811237}}.

\bibitem{Cremades:2003qj}
D.~Cremades, L.~E. Ibanez, and F.~Marchesano, ``Yukawa couplings in
  intersecting D-brane models,'' {\em JHEP} {\bf 07} (2003) 038,
\href{http://www.arXiv.org/abs/hep-th/0302105}{{\tt hep-th/0302105}}.

\bibitem{Douglas:2000ah}
M.~R. Douglas, B.~Fiol, and C.~Romelsberger, ``Stability and BPS branes,''
\href{http://www.arXiv.org/abs/hep-th/0002037}{{\tt hep-th/0002037}}.

\bibitem{Affleck:1983mk}
I.~Affleck, M.~Dine, and N.~Seiberg, ``Dynamical supersymmetry breaking in
  supersymmetric QCD,'' {\em Nucl. Phys.} {\bf B241} (1984)
493--534.

\bibitem{Blumenhagen:2005zh}
R.~Blumenhagen, G.~Honecker, and T.~Weigand, ``Non-abelian brane worlds: The
  open string story,''
\href{http://www.arXiv.org/abs/hep-th/0510050}{{\tt hep-th/0510050}}.

\bibitem{Douglas:2006xy}
M.~R. Douglas and W.~Taylor, ``The landscape of intersecting brane models,''
\href{http://www.arXiv.org/abs/hep-th/0606109}{{\tt hep-th/0606109}}.

\bibitem{Witten:1984dg}
E.~Witten, ``Some properties of O(32) superstrings,'' {\em Phys. Lett.} {\bf
  B149} (1984)
351--356.

\bibitem{Blumenhagen:2005pm}
R.~Blumenhagen, G.~Honecker, and T.~Weigand, ``Supersymmetric (non-)abelian
  bundles in the type I and SO(32) heterotic string,'' {\em JHEP} {\bf 08}
  (2005) 009,
\href{http://www.arXiv.org/abs/hep-th/0507041}{{\tt hep-th/0507041}}.

\bibitem{Blumenhagen:2005zg}
R.~Blumenhagen, G.~Honecker, and T.~Weigand, ``Non-abelian brane worlds: The
  heterotic string story,'' {\em JHEP} {\bf 10} (2005) 086,
\href{http://www.arXiv.org/abs/hep-th/0510049}{{\tt hep-th/0510049}}.

\bibitem{Blumenhagen:2005ga}
R.~Blumenhagen, G.~Honecker, and T.~Weigand, ``Loop-corrected compactifications
  of the heterotic string with line bundles,'' {\em JHEP} {\bf 06} (2005) 020,
\href{http://www.arXiv.org/abs/hep-th/0504232}{{\tt hep-th/0504232}}.

\bibitem{Silverstein:1995re}
E.~Silverstein and E.~Witten, ``Criteria for conformal invariance of (0,2)
  models,'' {\em Nucl. Phys.} {\bf B444} (1995) 161--190,
\href{http://www.arXiv.org/abs/hep-th/9503212}{{\tt hep-th/9503212}}.

\bibitem{Basu:2003bq}
A.~Basu and S.~Sethi, ``World-sheet stability of (0,2) linear sigma models,''
  {\em Phys. Rev.} {\bf D68} (2003) 025003,
\href{http://www.arXiv.org/abs/hep-th/0303066}{{\tt hep-th/0303066}}.

\bibitem{Beasley:2003fx}
C.~Beasley and E.~Witten, ``Residues and world-sheet instantons,'' {\em JHEP}
  {\bf 10} (2003) 065,
\href{http://www.arXiv.org/abs/hep-th/0304115}{{\tt hep-th/0304115}}.

\bibitem{Kachru:2003aw}
S.~Kachru, R.~Kallosh, A.~Linde, and S.~P. Trivedi, ``De Sitter vacua in string
  theory,'' {\em Phys. Rev.} {\bf D68} (2003) 046005,
\href{http://www.arXiv.org/abs/hep-th/0301240}{{\tt hep-th/0301240}}.

\bibitem{Achucarro:2006zf}
A.~Achucarro, B.~de~Carlos, J.~A. Casas, and L.~Doplicher, ``de Sitter vacua
  from uplifting D-terms in effective supergravities from realistic strings,''
  {\em JHEP} {\bf 06} (2006) 014,
\href{http://www.arXiv.org/abs/hep-th/0601190}{{\tt hep-th/0601190}}.

\bibitem{Burgess:2003ic}
C.~P. Burgess, R.~Kallosh, and F.~Quevedo, ``de Sitter string vacua from
  supersymmetric D-terms,'' {\em JHEP} {\bf 10} (2003) 056,
\href{http://www.arXiv.org/abs/hep-th/0309187}{{\tt hep-th/0309187}}.

\bibitem{Bergshoeff:2005yp}
E.~Bergshoeff, R.~Kallosh, A.-K. Kashani-Poor, D.~Sorokin, and A.~Tomasiello,
  ``An index for the Dirac operator on D3 branes with background fluxes,'' {\em
  JHEP} {\bf 10} (2005) 102,
\href{http://www.arXiv.org/abs/hep-th/0507069}{{\tt hep-th/0507069}}.

\bibitem{Lust:2005bd}
D.~L{\"u}st, P.~Mayr, S.~Reffert, and S.~Stieberger, ``F-theory flux,
  destabilization of orientifolds and soft terms on D7-branes,'' {\em Nucl.
  Phys.} {\bf B732} (2006) 243--290,
\href{http://www.arXiv.org/abs/hep-th/0501139}{{\tt hep-th/0501139}}.

\bibitem{Lust:2005cu}
D.~L{\"u}st, S.~Reffert, W.~Schulgin, and P.~K. Tripathy, ``Fermion zero modes
  in the presence of fluxes and a non- perturbative superpotential,''
\href{http://www.arXiv.org/abs/hep-th/0509082}{{\tt hep-th/0509082}}.

\bibitem{Baumann:2006th}
D.~Baumann {\em et al.}, ``On D3-brane potentials in compactifications with
  fluxes and wrapped D-branes,''
\href{http://www.arXiv.org/abs/hep-th/0607050}{{\tt hep-th/0607050}}.

\bibitem{Cvetic:2001nr}
M.~Cveti{\v c}, G.~Shiu, and A.~M. Uranga, ``Chiral four-dimensional N = 1
  supersymmetric type IIA orientifolds from intersecting D6-branes,'' {\em
  Nucl. Phys.} {\bf B615} (2001) 3--32,
\href{http://www.arXiv.org/abs/hep-th/0107166}{{\tt hep-th/0107166}}.

\bibitem{Cvetic:2001tj}
M.~Cveti{\v c}, G.~Shiu, and A.~M. Uranga, ``Three-family supersymmetric
  standard like models from intersecting brane worlds,'' {\em Phys. Rev. Lett.}
  {\bf 87} (2001) 201801,
\href{http://www.arXiv.org/abs/hep-th/0107143}{{\tt hep-th/0107143}}.

\bibitem{Cremades:2002va}
D.~Cremades, L.~E. Ibanez, and F.~Marchesano, ``Towards a theory of quark
  masses, mixings and CP- violation,''
\href{http://www.arXiv.org/abs/hep-ph/0212064}{{\tt hep-ph/0212064}}.

\bibitem{Cvetic:2004ui}
M.~Cveti{\v c}, T.~Li, and T.~Liu, ``Supersymmetric Pati-Salam models from
  intersecting D6- branes: A road to the standard model,'' {\em Nucl. Phys.}
  {\bf B698} (2004) 163--201,
\href{http://www.arXiv.org/abs/hep-th/0403061}{{\tt hep-th/0403061}}.

\bibitem{Nath:2006ut}
P.~Nath and P.~F. Perez, ``Proton stability in grand unified theories, in
  strings, and in branes,''
\href{http://www.arXiv.org/abs/hep-ph/0601023}{{\tt hep-ph/0601023}}.

\bibitem{Allanach:2003eb}
B.~C. Allanach, A.~Dedes, and H.~K. Dreiner, ``The R parity violating minimal
  supergravity model,'' {\em Phys. Rev.} {\bf D69} (2004) 115002,
\href{http://www.arXiv.org/abs/hep-ph/0309196}{{\tt hep-ph/0309196}}.

\bibitem{CveticLangacker}
M.~Cveti{\v c} and P.~Langacker, ``New grand unified models with intersecting
  D6-branes, neutrino masses, and flipped SU(5),''
\href{http://www.arXiv.org/abs/hep-th/0607238}{{\tt hep-th/0607238}}.

\bibitem{Blumenhagen:2004xx}
R.~Blumenhagen, F.~Gmeiner, G.~Honecker, D.~Lust, and T.~Weigand, ``The
  statistics of supersymmetric D-brane models,'' {\em Nucl. Phys.} {\bf B713}
  (2005) 83--135,
\href{http://www.arXiv.org/abs/hep-th/0411173}{{\tt hep-th/0411173}}.

\bibitem{Gmeiner:2005vz}
F.~Gmeiner, R.~Blumenhagen, G.~Honecker, D.~Lust, and T.~Weigand, ``One in a
  billion: MSSM-like D-brane statistics,'' {\em JHEP} {\bf 01} (2006) 004,
\href{http://www.arXiv.org/abs/hep-th/0510170}{{\tt hep-th/0510170}}.

\bibitem{Dijkstra:2004cc}
T.~P.~T. Dijkstra, L.~R. Huiszoon, and A.~N. Schellekens, ``Supersymmetric
  standard model spectra from RCFT orientifolds,'' {\em Nucl. Phys.} {\bf B710}
  (2005) 3--57,
\href{http://www.arXiv.org/abs/hep-th/0411129}{{\tt hep-th/0411129}}.

\bibitem{Anastasopoulos:2006da}
P.~Anastasopoulos, T.~P.~T. Dijkstra, E.~Kiritsis, and A.~N. Schellekens,
  ``Orientifolds, hypercharge embeddings and the standard model,''
\href{http://www.arXiv.org/abs/hep-th/0605226}{{\tt hep-th/0605226}}.

\end{thebibliography}\endgroup
\bibliographystyle{utphys}

\end{document}